\documentclass[12pt]{elsart}

\usepackage[thinlines,thiklines]{easytable}
\usepackage[thinlines,thiklines]{easymat}
\usepackage{array}
\usepackage{amsmath,amsthm}
\usepackage{amsfonts}
\usepackage[dvips]{graphicx}
\usepackage{epsfig}
\usepackage{amssymb}
\usepackage{latexsym}
\usepackage{multirow}

\def\etal{et al.}

\newcommand{\mrom}[1]{\ensuremath{\mathrm{#1}}}
\newcommand{\deesix}{\ensuremath{\mrom{D}_6}}
\newcommand{\nom}[1]{\ensuremath{|#1|}}
\newcommand{\nomsq}[1]{\ensuremath{{|#1|}^2}}

\newcommand{\ud}{\ensuremath{\mathrm{d}}}
\newcommand{\rombf}[1]{\ensuremath{\mathbf{\mrom{#1}}}}

\newcommand{\haf}{\ensuremath{\frac{1}{2}}}

\newcommand{\ie}{i.e.,\/ }

\newcommand{\dv}{\mathbf{d}}
\newcommand{\ev}{\mathbf{e}}

\newcommand{\gv}{\mathbf{g}}
\newcommand{\hv}{\mathbf{h}}
\newcommand{\lv}{\mathbf{l}}

\newcommand{\xv}{\mathbf{x}}

\newcommand{\zv}{\mathbf{z}}
\newcommand{\kv}{\mathbf{k}}
\newcommand{\Kv}{\mathbf{K}}

\newcommand{\ro}{\ensuremath{\rho}}
\newcommand{\rosixty}{\ro}

\newcommand{\ton}{\ensuremath{\tau_{1}}}
\newcommand{\ttw}{\ensuremath{\tau_{2}}}
\newcommand{\toc}{\ensuremath{\tau_{1}^3}}
\newcommand{\ttc}{\ensuremath{\tau_{2}^3}}

\begin{document}
\begin{frontmatter}

\title{Spatial period-multiplying instabilities of hexagonal Faraday waves}

\author{D.P.~Tse\thanksref{DAWNADDRESS}}, 
\author{A.M.~Rucklidge\thanksref{ALASTAIREMAIL}}, 
\author{R.B.~Hoyle}
\address{Department of Applied Mathematics and Theoretical Physics,\\
University of Cambridge, Silver Street, Cambridge CB3 9EW, UK}
\and
\author{M.~Silber\thanksref{MARYEMAIL}}
\address{Department of Engineering Sciences and Applied Mathematics,\\
Northwestern University, Evanston, IL 60208 USA}

 \thanks[DAWNADDRESS]{Current address: Warburg Dillon Read, 100 Liverpool 
Street, London EC2M 2RH, UK}
 \thanks[ALASTAIREMAIL]{E-mail: A.M.Rucklidge@damtp.cam.ac.uk}
 \thanks[MARYEMAIL]{E-mail: m-silber@northwestern.edu}

\begin{abstract}
A recent Faraday wave experiment with two-frequency forcing reports two types
of `superlattice' patterns that display periodic spatial structures having two
separate scales~\cite{KudrPier98}. These patterns both arise as secondary
states once the primary hexagonal pattern becomes unstable. In one of these
patterns (so-called `superlattice-two') the original hexagonal symmetry is
broken in a subharmonic instability to form a striped pattern with a spatial
scale increased by a factor of $2\sqrt{3}$ from the original scale of the
hexagons. In contrast, the time-averaged pattern is periodic on a hexagonal
lattice with an intermediate spatial scale ($\sqrt{3}$~larger than the original
scale) and apparently has $60^{\circ}$~rotation symmetry. We present a
symmetry-based approach to the analysis of this bifurcation. Taking as our
starting point only the observed instantaneous symmetry of the superlattice-two
pattern presented in~\cite{KudrPier98} and the subharmonic nature of the
secondary instability, we show (a)~that a pattern with the same instantaneous
symmetries as the superlattice-two pattern can bifurcate stably from standing
hexagons; (b)~that the pattern has a spatio-temporal symmetry not reported
in~\cite{KudrPier98}; and (c)~that this spatio-temporal symmetry accounts for
the intermediate spatial scale and hexagonal periodicity of the time-averaged
pattern, but not for the apparent $60^{\circ}$~rotation symmetry. The approach
is based on general techniques that are readily applied to other secondary
instabilities of symmetric patterns, and does not rely on the primary pattern
having small amplitude.
 \end{abstract}


\end{frontmatter}

\section{Introduction}\label{sec:intro}

The classical hydrodynamic problem of parametrically driven surface waves -- or
Faraday waves -- concerns the spontaneous generation of standing waves at the
free surface of a horizontal layer of fluid when subjected to vertical
oscillations whose amplitude exceeds a critical value. Its usefulness as a tool
to study nonlinear pattern-forming dynamics in non-equilibrium systems is
reflected in the considerable amount of interest shown in the subject by
experimentalists and theoreticians alike. A review of earlier works, mostly
conducted with low-viscosity fluids in small vessels and a single forcing
frequency, can be found in~\cite{MilesH90}. More recently, Edwards \&
Fauve~\cite{EdwaFauv94} have performed experiments in the small-depth,
high-viscosity and large-aspect ratio regime using a forcing function with two
commensurate frequency components that modulates gravity periodically. In this
regime, where it can be shown that the wavenumber of the selected pattern is
less sensitive to the size and shape of the container, and that long-wavelength
modes are heavily damped, observations of spatially periodic patterns (stripes,
squares, hexagons), circular patterns (targets and spirals) and quasi-patterns
have been reported~\cite{EdwaFauv94,KudrolliG96}. A survey of more recent
results has been carried out by M\"{u}ller \etal~\cite{MullerFP98}. Over the
past two years, a new class of `superlattice patterns' has been independently
observed by Kudrolli \etal~\cite{KudrPier98} and Arbell \&
Fineberg~\cite{ArbellF98} in experiments employing two-frequency forcing
functions, and by Wagner \etal~\cite{WagnerMK98} in experiments using
non-Newtonian fluids. These superlattices are so termed because of their
distinctive feature of having spatial structures on two different length scales
when viewed at any instant in time~\cite{KudrPier98}. Steady patterns that
display similar characteristics have also been observed in convection
experiments on fluids with temperature-dependent viscosity~\cite{White88} and
have been investigated in a model of long wave convection~\cite{SkeldonS98} and
in reaction-diffusion systems near a Turing bifurcation~\cite{JuddS98}.

Two types of superlattice patterns have been reported in~\cite{KudrPier98} for
different parameter values. Both of them, despite their different spatial and
spatio-temporal symmetry properties, are found to be possible transitions from
harmonic standing hexagons as the forcing amplitude is increased. (In this
context, {\em harmonic} indicates an oscillation with the same period as that
of the external forcing, denoted by~$T$, while {\em subharmonic} indicates an
oscillation with twice that period.) The first of these patterns (called
`superlattice-one' by Kudrolli \etal~\cite{KudrPier98}) is a harmonic response
with triangular symmetry on a small scale and hexagonal lattice periodicity on
a larger scale. This pattern has been studied by Silber \&
Proctor~\cite{SilberP98}, who showed that it (along with standing hexagons) can
arise in a bifurcation from the flat, undisturbed state when a hexagonal
lattice with spatial periodicity larger than that dictated by the critical
wavelength is considered~\cite{DionneG92,DionSkel97}. Silber \&
Proctor~\cite{SilberP98} also suggested that stability might be transfered from
standing hexagons to superlattice-one through an intermediate branch.

\begin{center}
\begin{figure}[htbp]
 \hfil (a) \hfil (b) \hfil (c) \\
 \smallskip
  {\resizebox{!}{6.26truecm}{\includegraphics{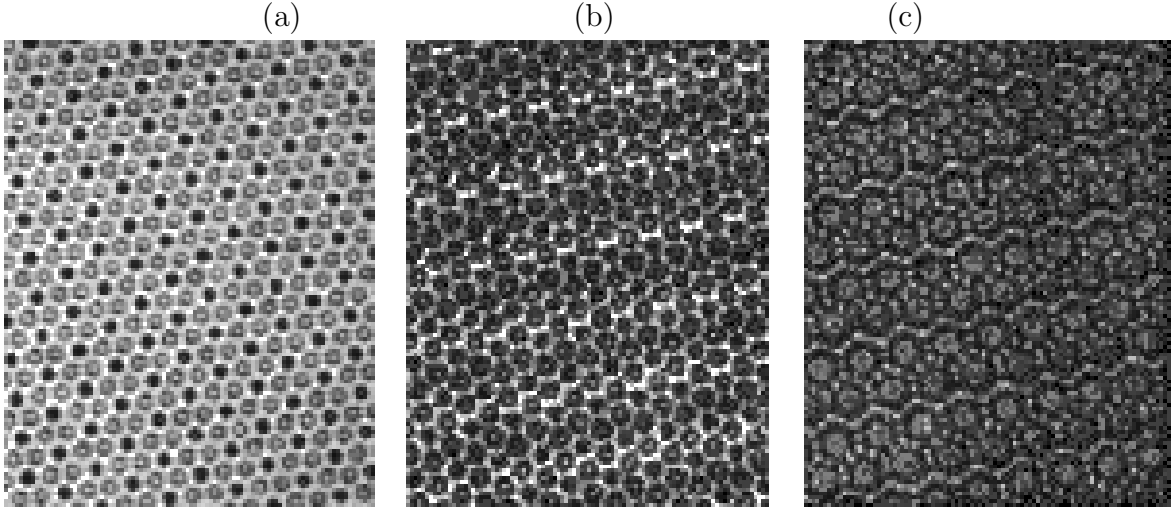}}}\\
  \caption{\sl \small (a) Time-averaged image of the superlattice-two pattern
displays a well-defined hexagonal symmetry on two spatial scales. (b) \& (c)
Instantaneous snapshots of the same pattern separated by $3/20$th of the
external driving period $T$ reveals a time-dependent stripe-like modulation.
The pattern in (a) is a different realization of the experiment from those in
(b) \&~(c). Courtesy of Kudrolli \etal~\cite{KudrPier98}, reproduced with
permission.}
 \label{fig:original}  
 \end{figure}
 \end{center}

The second type of superlattice pattern (`superlattice-two'~\cite{KudrPier98}),
in contrast to the first, arises in a period-doubling (or subharmonic)
instability of the standing hexagons. If we let $u(\xv,t)$ measure the
deformation of the free fluid surface at time~$t$, it satisfies
 \begin{equation}\label{pdouble}
 u(\xv,t+2T) = u(\xv,t), \quad
 u(\xv,t+T) \neq u(\xv,t). 
 \end{equation}
Further, this pattern exhibits a complicated mixture of spatial symmetry and
time-dependent behaviour. When averaged over two periods of the driving
function, its image displays hexagonal symmetry with two well-defined spatial
scales in the ratio $1:\sqrt{3}$ (\mbox{figure~\ref{fig:original}(a)}).
Remarkably, at any instant, a wavy, stripe-like spatial modulation destroys the
average hexagonal symmetry, resulting in a pattern that appears vastly
different from its time-averaged image (see figures~\ref{fig:original}(b)
and~\ref{fig:original}(c)). Arbell \& Fineberg (unpublished) have also found
the superlattice-two pattern for similar experimental parameters.

The superlattice-two pattern presents a number of theoretical challenges that
motivate this paper: the disappearance of the stripes from the time-averaged
pattern; the reduced spatial period in the time-average; and the apparent
$60^{\circ}$~rotation symmetry of the time-average. We present a symmetry-based
approach to the study of this pattern by taking the view that it arises as a
symmetry-breaking instability from the underlying standing hexagons in a
spatial period-multiplying bifurcation. Our aim is to classify qualitatively
the range of possible bifurcating solutions and to understand how their
symmetry properties can be related to the experimental observations described
above. We emphasize that we are examining instabilities of fully nonlinear
states, so our approach differs from weakly nonlinear studies of the primary
Faraday instability.

There are three stages in our approach. First, by using the experimentally
observed instantaneous spatial symmetry information of the superlattice-two
instability and by making the assumption that all solutions are periodic in the
plane, we can restrict all patterns to a suitably chosen spatially periodic
lattice. This lattice in turn defines a compact symmetry group, which we denote
by~$\Gamma^s$, with the key properties that its action leaves standing hexagons
invariant and that it has a subgroup, which we denote by~$\Sigma^s$, that
describes the {\em instantaneous} spatial symmetry of the bifurcating
superlattice pattern. Due to the compactness and special structure
of~$\Gamma^s$, we can compute explicitly all its irreducible representations.
Second, we observe that since $\Sigma^s$ is by definition the isotropy subgroup
of the bifurcating solution under the action of~$\Gamma^s$, it must have a
non-trivial fixed-point set. This restriction allows us to identify the one
relevant irreducible representation of $\Gamma^s$ that describes the spatial
symmetry properties of the marginal eigenfunctions at the superlattice
bifurcation point. Finally, by considering the action of the time-shift
symmetry $\tau_t: t \rightarrow t + T$  on the period-doubling marginal modes,
we obtain the irreducible representation of the full symmetry group (denoted
by~$\Gamma$) and hence the normal form of the bifurcation problem. We can then
invoke the equivariant branching lemma~\cite{GolubitskySS88} to show that there
are at least six primary branches of solutions bifurcating from standing
hexagons.

With one proviso, the superlattice-two pattern observed by
Kudrolli~\etal~\cite{KudrPier98} can be identified as one of these branches,
which we show can bifurcate as a stable branch from standing hexagons. By
applying techniques for studying the averaged symmetry of periodic orbits
(cf~\cite{BaranyDG93}), we show that the time-average of this branch of
solutions has the hexagonal lattice periodicity observed in the experiment (as
in figure~\ref{fig:original}(a)); this change in the spatial length scale on
time-averaging is a consequence of the branch of solutions possessing a
spatio-temporal symmetry. This symmetry was not reported in~\cite{KudrPier98}.
The proviso mentioned above is that our time-average pattern does not possess
$60^{\circ}$~rotation symmetry -- we will return to this discrepancy in the
final section.

A further stability analysis predicts that other patterns, displaying different
spatial and spatio-temporal symmetry properties, can bifurcate as stable
branches of solutions from standing hexagons in different regions of parameter
space. More generally, our analysis indicates that patterns that display
superlattice structures can arise in two-dimensional spatial period-multiplying
bifurcations from an underlying non-trivial solution, and our approach could
also be used to investigate other superlattice patterns of Arbell \&
Fineberg~\cite{ArbellF98} and Wagner \etal~\cite{WagnerMK98}. In particular, it
may be possible to analyse some of those experimental results in terms of other
irreducible representations of the same group~$\Gamma^s$.

The issue of spatial period-multiplying instabilities is an interesting one
that has arisen in a variety of experimental and theoretical contexts.
Period-multiplying bifurcations in one lateral direction have arisen in
convection problems~\cite{McKenzie88}, magnetoconvection~\cite{ProctorWeiss93},
Taylor--Couette experiments~\cite{Iooss86} and in numerical solutions of the
Kuramoto--Sivashinsky equations~\cite{AstonSW92,AmdjadiAP97}. Much less is
known about spatial period-multiplying bifurcations in two directions. There
are now several experimental observations of this phenomenon in the Faraday
wave problem~\cite{KudrPier98,ArbellF98,WagnerMK98} as well as in convection
experiments~\cite{White88,McKenzie88,ProcMatt96} and magnetoconvection
calculations~\cite{Dawes2000,RWBMP2000}.

In the next section, we introduce some fundamental definitions and results from
equivariant bifurcation theory~\cite{GolubitskySS88} to help us describe how
this problem can be cast into a theoretical framework. In
section~\ref{sec:findgamma}, we fully describe the symmetry group of the
bifurcation problem that will give rise to the observed symmetry-breaking
behaviour. We also show that, under suitable phenomenological assumptions, we
can identify and hence explicitly compute the irreducible representation that
is relevant to the action of the symmetry group on the observed bifurcating
modes. The normal form of the bifurcation problem and a stability analysis are
presented in section~\ref{sec:bifprob}. Discussions of our approach and a
comparison with the experiments follow in section~\ref{sec:discuss}.

\section{Group theoretic ideas}\label{sec:groupideas}

In order to study the superlattice-two pattern as a symmetry-breaking
instability from standing hexagons, it is necessary to identify all the
symmetries that are initially present. Due to the apparent absence of side-wall
effects in the observed patterns, we consider the mathematical idealisation
that all physical fields are defined in a laterally unbounded domain. Standing
hexagons are then easily seen to be invariant under the action generated by a
reflection, a $60^{\circ}$~rotation, and two linearly independent translations.
The group generated by these symmetry actions is isomorphic to $\Zset^2
\dotplus \deesix$, which is non-compact. (Here $\Zset$ denotes the group of
integers under addition, and $\deesix$ is the twelve-element symmetry group of
a regular hexagon.) Consequently a bifurcation problem that is equivariant
under the action of this group can have an infinite number of modes related by
symmetry becoming marginally stable simultaneously. This difficulty can be
resolved if we restrict possible solutions to doubly-periodic functions defined
on a suitably chosen lattice, an assumption justified by the distinct spatial
periodicity of the observed patterns. A suitable lattice, which can be viewed
as a finite cell with periodic boundary conditions, is one that captures the
spatial periodicity of both the bifurcating modes as well as the standing
hexagons. With respect to such a periodic cell, the symmetries that leave
standing hexagons invariant now form a finite, and hence compact group that can
be studied via representation theory. Our task therefore, is to make use of the
available symmetry information taken from experimental observations to choose a
lattice on which we can define a suitable spatial symmetry group $\Gamma^s$
with the properties outlined in section~\ref{sec:intro}. The idea of
suitability can be made precise after we have introduced some basic group
theoretic results.

Since we are considering bifurcations from a time-periodic solution, we
formulate the bifurcation problem of the superlattice-two pattern, a
period-doubling instability, by expanding about standing hexagons using a
stroboscopic map $\mathcal{G}$ in the manner described by Crawford \&
Knobloch~\cite{CrawKnob91} and Silber \& Proctor~\cite{SilberP98}.
Specifically, we are assuming that standing hexagons, a fixed point of the
map~$\mathcal{G}$, lose stability to subharmonic waves with period $2T$ as a
bifurcation parameter $\mu$ is varied past zero. This implies that the
linearised map $D\mathcal{G}$ evaluated at the fixed point has a real
eigenvalue passing through the value $-1$ as the bifurcating waves become
unstable. With the assumption that all fields are defined in a periodic cell
such that symmetries of the standing hexagons are described by a compact
group~$\Gamma^s$, the linearised map $D\mathcal{G}$ has a finite number ($p$)
of marginal eigenfunctions associated with the eigenvalue $-1$ as $\mu$ crosses
zero. We denote the amplitudes of these $p$ marginal modes at time $t=qT, \; q
\in \Zset$ by $\zv_q = \left[z_1(qT), \ldots , z_p(qT)\right]\in\Rset^p$. In
addition, the pattern has two neutrally stable modes (eigenvalues equal to~$1$)
associated with translations of the standing hexagons
(see~\cite{Iooss86,RuckSilb98}); the amplitudes of these two modes, which
correspond to the translation of the pattern in the plane, are denoted
by~$\dv_q$. Close to the onset of the period-doubling instability,
$\mathcal{G}$~can be reduced to a finite-dimensional map $\gv$ defined on the
centre manifold spanned by these amplitudes:
 \begin{equation}\label{normalform}
 \zv_{q+1} = \gv(\zv_q; \mu), \quad 
 \gv: \Rset^p \times \Rset \rightarrow \Rset^p,
 \end{equation}
coupled with a map $\hv: \Rset^p \times \Rset \rightarrow \Rset^2$ describing
how the perturbation drives translations of the pattern:
 \begin{equation}\label{drifteqn}
 \dv_{q+1} = \dv_q + \hv(\zv_q;\mu).
 \end{equation}
The map $\gv$ is forced by symmetry to be $\Gamma^s$-equivariant:
 \begin{equation}\label{gammaequiv}
 \gamma \gv(\zv_q;\mu) = \gv(\gamma \zv_q;\mu) \quad \mbox{for all}\; 
 \gamma \in \Gamma^s,
 \end{equation}
while the map $\hv$ obeys
 \begin{equation}\label{hequivariant}
 N_{\gamma}\hv\left(\zv_q;\mu\right) = 
 \hv\left(\gamma \zv_q; \mu\right) \quad \mbox{for all}\;
 \gamma \in \Gamma^s,
 \end{equation}
where $N_\gamma$ is the $2\times 2$ matrix that represents how the symmetry
$\gamma$ acts on a horizontal displacement vector~\cite{RuckSilb98}. In terms
of the marginal modes, standing hexagons correspond to the trivial state,
\mbox{$\zv = \mathbf{0}$.} When considered as a symmetry-breaking bifurcation
from the underlying standing hexagons, the superlattice pattern corresponds to
a non-trivial, period-two solution to the map~$\gv$, denoted
by~$\zv^{\ast}_{q}$, whose instantaneous spatial symmetry is specified by its
isotropy subgroup $\Sigma_{\zv^{\ast}_{q}}^s$:
 \begin{equation}\label{isotropy}
 \Sigma_{\zv^{\ast}_{q}}^s = \left\{\sigma \in \Gamma^s: \sigma 
                         \zv^{\ast}_{q} = \zv^{\ast}_q \right \} 
                         \subset \Gamma^s.
 \end{equation}
In fact, this solution must lie in the fixed-point subspace of
$\Sigma_{\zv^{\ast}_q}^s$:
 \begin{equation}\label{fixsigma}
 {\mbox{Fix}}(\Sigma_{\zv^{\ast}_q}^s) = \left\{\zv \in \Rset^p : 
 \sigma \zv = \zv, \;\mbox{for all}\; 
 \sigma \in \Sigma_{\zv^{\ast}_q}^s\right\},
 \end{equation}
which is a linear subspace of $\Rset^p$ and invariant
under~$\gv$~\cite{GolubitskySS88}.

Since standing hexagons do not possess spatio-temporal
symmetries~\cite{RobertsSW86}, $\Gamma^s$-equivariance is sufficient to
determine the normal form of the map $\gv$ in the case of (temporal)
period-preserving bifurcations. However, for period-doubling bifurcations there
is an extra symmetry pertaining to the normal form, related to the time-shift
action $\tau_t: t \rightarrow t + T$ on the bifurcating modes. In this case,
the normal form of the map $\gv$ is $\left(\Gamma^s \times \mrom{Z}_2
\right)$-equivariant~\cite{ChossatG88,LambM99}. Once the full symmetry group of
the normal form of $\gv$ is determined, we can apply the equivariant branching
lemma, which, with suitable interpretation, states that if certain
non-degeneracy conditions are satisfied, there is a unique branch of
bifurcating solutions for each isotropy subgroup of $\Gamma \equiv \Gamma^s
\times \mrom{Z}_2$ with a one-dimensional fixed-point subspace. So instead of
solving for solutions of the nonlinear vector field~$\gv$, we can simply look
for isotropy subgroups of $\Gamma$ with this property. To apply the equivariant
branching lemma, we need to know explicitly how symmetry acts on all the
marginal modes, but experimental observations only provide information about
the instantaneous spatial symmetry of one of these modes. We cannot infer
directly from the observations the total number of marginal modes that are
related by symmetry at the bifurcation point, nor the set of matrices that
represent the action of the symmetry group $\Gamma$ on the marginal modes and
the map~$\gv$. However, this difficulty can be resolved if we make the
(generic) assumption that the bifurcation is associated with an irreducible
representation of the group~$\Gamma$, and this is where the need to invoke
representation theory arises.

In order to introduce the key properties of irreducible representations
(irreps) and describe how they can be computed for a finite group~$\Gamma$, we
recall the following definitions~\cite{Cornwell}.
 \begin{enumerate}
 \renewcommand{\labelenumi}{(\roman{enumi})}
 \item A {\em representation} of the group $\Gamma$ is a homomorphism $\psi$
that maps $\Gamma$ into a set of invertible $n\times n$ matrices
$\rombf{M}_{\Gamma}$ acting on $\Rset^n$ or $\Cset^n$, in other words
 \begin{displaymath}
 \psi\left(\gamma\right) = M_{\gamma}, \quad \quad
 \gamma \in \Gamma,\; M_{\gamma} \in \rombf{M}_{\Gamma}
 \end{displaymath}
such that $\psi\left(\gamma_1 \gamma_2\right) = \psi(\gamma_1)\psi(\gamma_2)\:
\mbox{for all}\, \gamma_1, \gamma_2 \in \Gamma$. The integer~$n$ is the {\em
dimension} of the representation.
 \item Two $n$-dimensional representations $\rombf{M}_{\Gamma}$ and
$\rombf{N}_{\Gamma}$ of $\Gamma$ are called {\em equivalent} if there is an
invertible $n \times n$ matrix $Q$ such that for each $\gamma \in \Gamma$,
 \begin{displaymath}
 N_\gamma = Q^{-1}M_\gamma Q, \quad M_\gamma \in \rombf{M}_{\Gamma},\,
 N_\gamma \in \rombf{N}_{\Gamma}. 
 \end{displaymath}
 \item A {\em conjugacy class} of $\Gamma$ is a subset $C$ of $\Gamma$ such
that $\gamma^{-1}c\gamma \in C, \mbox{for all}\, c \in C \;\mbox{and}\;
\gamma\in\Gamma$.
 \item The {\em character} of an element $\gamma \in \Gamma$ in a
representation $\rombf{M}_{\Gamma}$ is defined to be the trace of the matrix
$M_{\gamma}$, and we denote this value by~$\chi_{M_{\gamma}}$.
 \item A representation of $\Gamma$ on $\Rset^n$ ($\Cset^n$) is said to be {\em
irreducible} if it does not leave invariant any proper subspace of $\Rset^n$
($\Cset^n$).
 \end{enumerate}
 Simplistically we can consider a representation as a set of $n\times n$
nonsingular matrices that specifies the action of $\Gamma$ on the vector space
$\Rset^n$ or $\Cset^n$ and at the same time preserves the group structure. It
is possible to show that every representation of a finite group is equivalent
to a unitary representation -- one in which all matrices are
unitary~\cite{Cornwell}. A simple result of definition (iv) is that the
character of the identity element in a representation is always equal to the
dimension~$n$ of that representation, and definitions (i), (iii) and (iv) imply
that elements in the same conjugacy class have the same character. The
characters of the irreps of $\Gamma$ obey a set of rules inherited from the
orthogonality theorem governing the underlying irreps~\cite{RileyHB98} and for
simple groups such as $\mrom{Z}_2$, $\mrom{Z}_6$ and $\deesix$, the character
tables can easily be constructed by appealing to those rules. The orthogonality
theorem also implies that the number of irreps of a group is equal to the
number of conjugacy classes. For finite groups with a semi-direct product
structure of the form $\Gamma = \mathcal{A} \dotplus \mathcal{B}$ such that
$\mathcal{A}$ is a normal (or invariant) subgroup of~$\Gamma$ (that is,
$\gamma^{-1}a\gamma\in\mathcal{A}$ for every $a\in\mathcal{A}$ and
$\gamma\in\Gamma$), the characters of the irreps of $\mathcal{A}$ and
$\mathcal{B}$ form the building blocks in determining all the characters and
constructing unitary irreps of the group $\Gamma$ via a special
algorithm~\cite{Cornwell}.

In summary, analysis of the superlattice pattern using these group theoretic
tools depends on our being able to find a spatial lattice or periodic cell on
which standing hexagons and the marginal modes exhibiting the observed
symmetries fit. The arrangement of the standing hexagons in the periodic cell
then gives us a suitable symmetry group~$\Gamma^s$, which has a subgroup
$\Sigma_{\zv^{\ast}_q}^s$, defined in~\eqref{isotropy}, whose elements are
determined from experimental observations. Once we have calculated all the
characters of~$\Gamma^s$, the restriction provided by the requirement that
$\Sigma_{\zv^{\ast}_q}^s$ be the isotropy subgroup of the observed pattern
enables us to isolate the one irrep that describes the action of $\Gamma^s$ on
all the marginal modes related to the observed bifurcating mode by symmetry.
Indeterminacy in the choice of irreps can be avoided if we choose a unit cell
that captures exactly one spatial period of the observed pattern. The details
of this procedure are the subject of the next section.

 \begin{center}
 \begin{figure}[htbp]
  \begin{tabular}{ccc}
    (a) & (b) & (c) \\ 
    \resizebox{!}{6.2cm}{\includegraphics{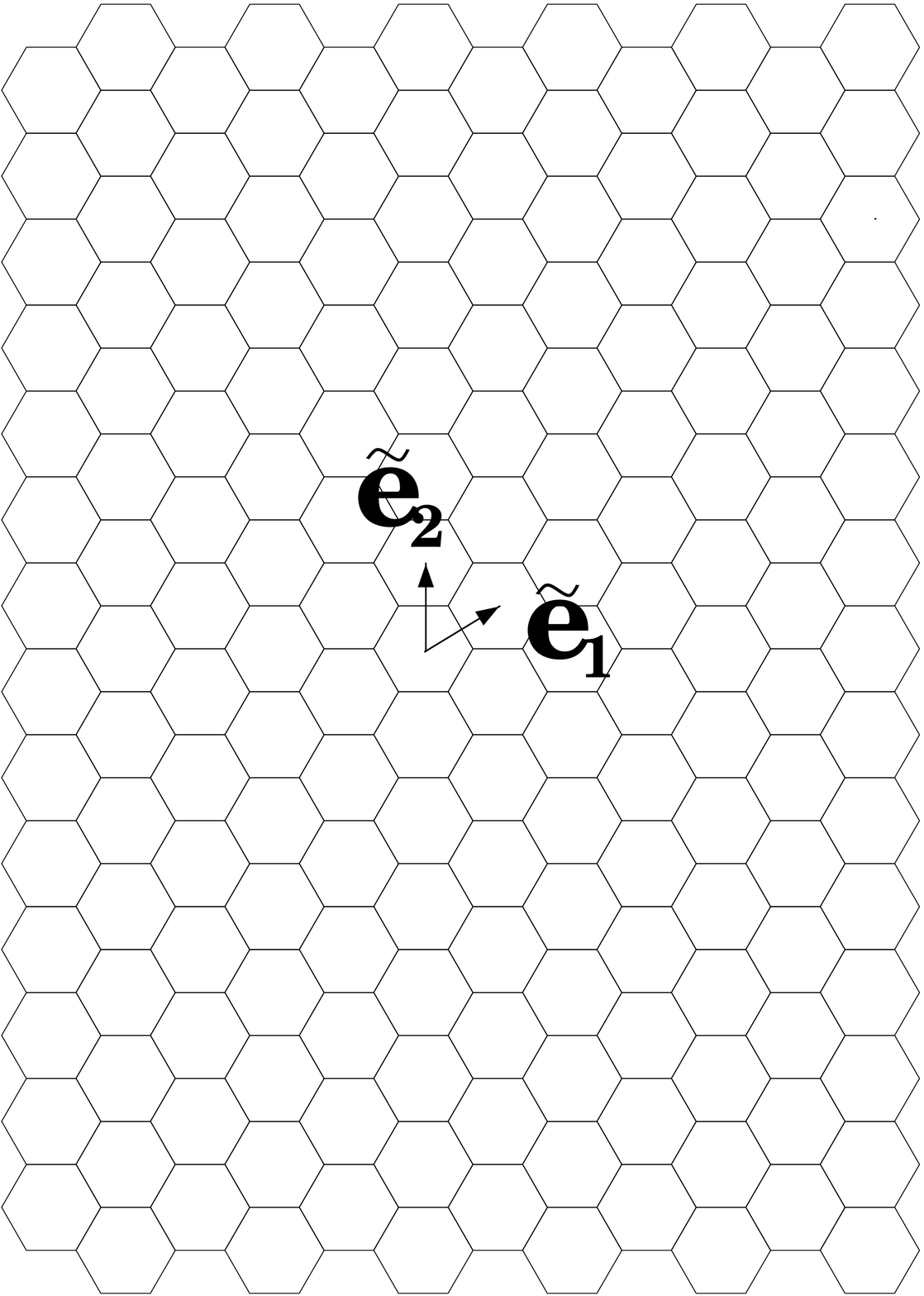}} &
    \resizebox{!}{6.2cm}{\includegraphics{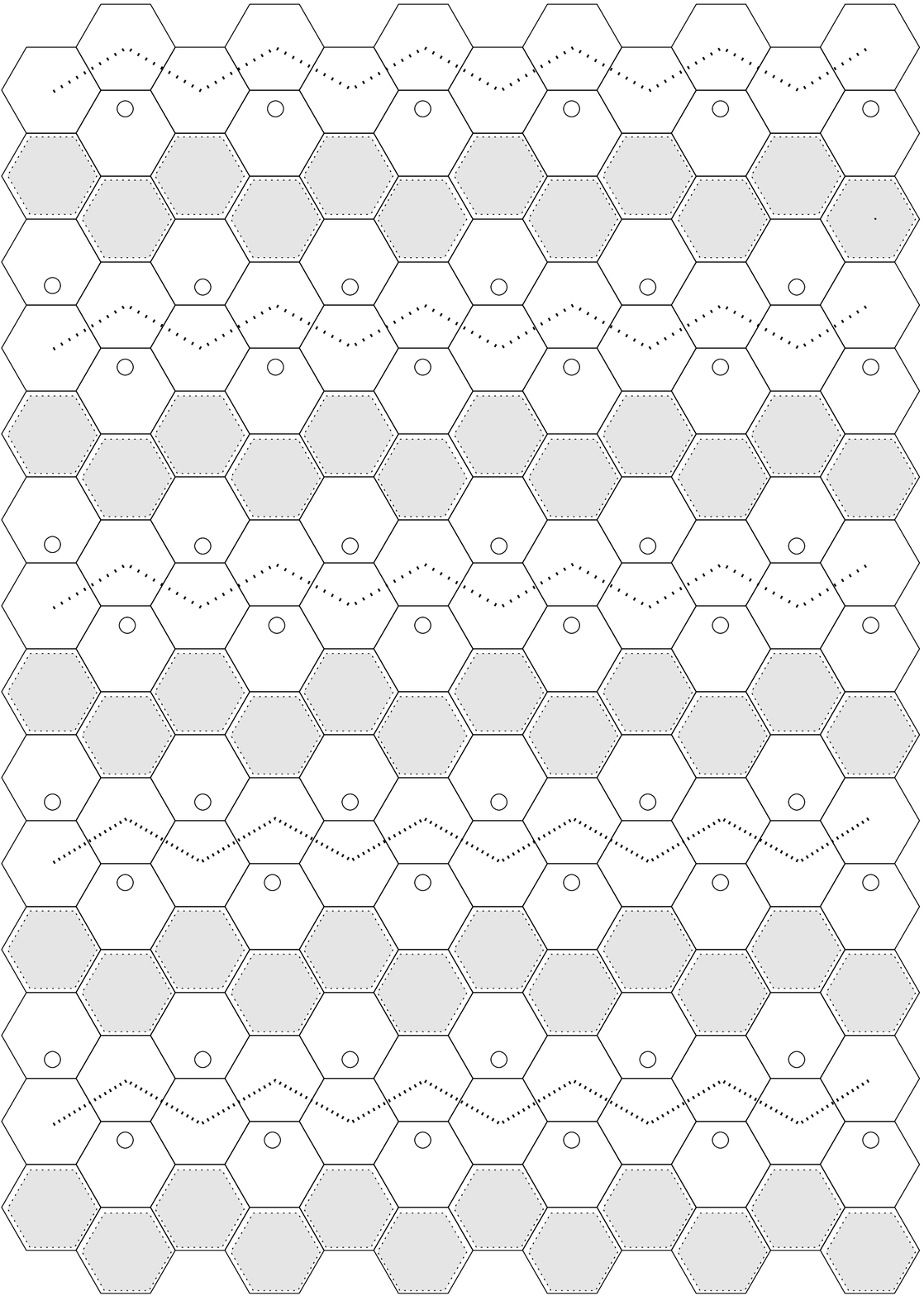}} &
    \resizebox{!}{6.2cm}{\includegraphics{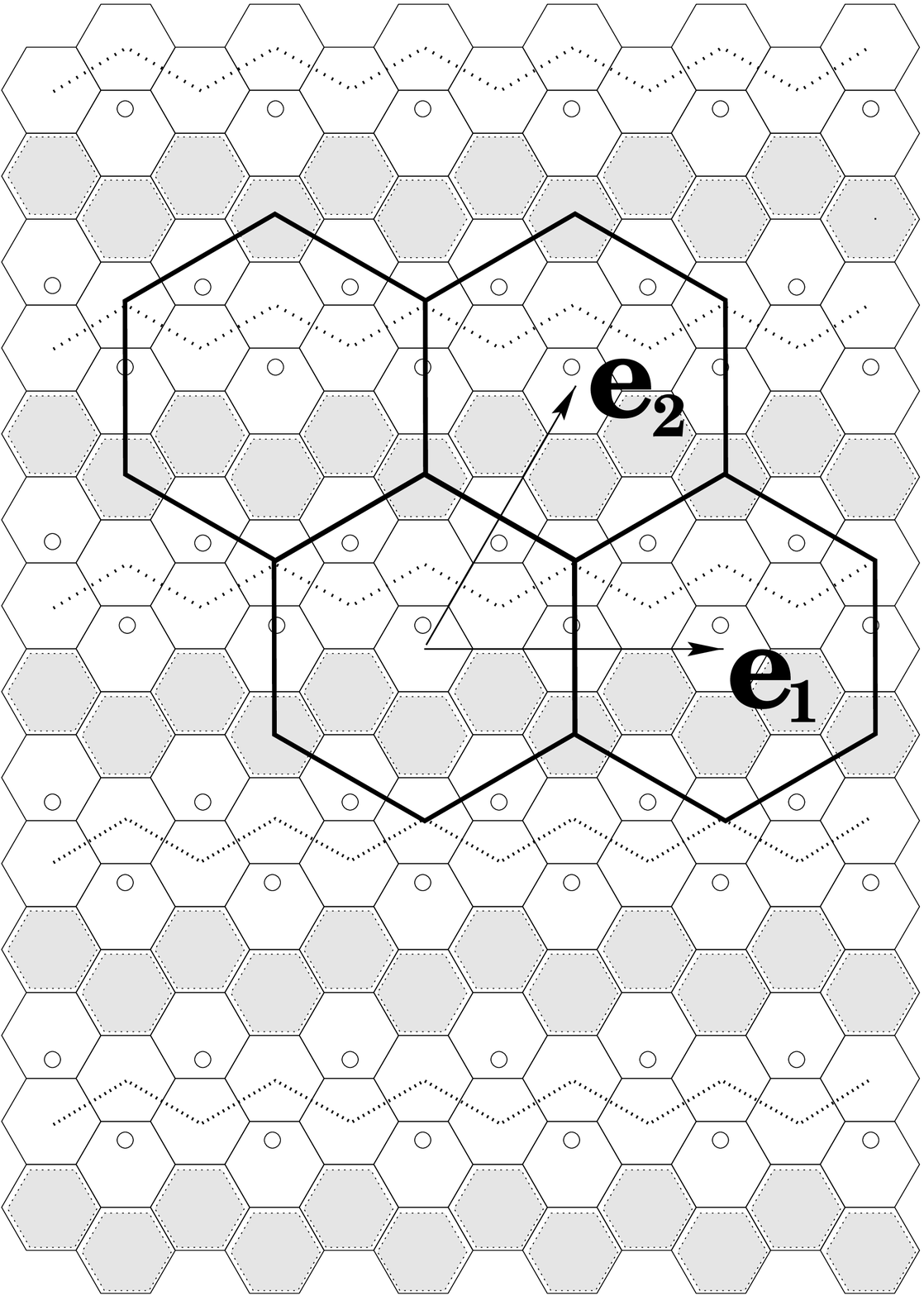}} \\
  \end{tabular}
 \caption{\sl \small (a) Schematic representation of standing hexagons, where
$\tilde{\ev}_1$ and $\tilde{\ev}_2$ denote the vectors of translations defined
in~(\ref{tauonetwo}). (b)~Diagram depicting the instantaneous spatial symmetry
of the superlattice-two pattern (see figure~\ref{fig:original}(c)), whose
spatial periodicity can be captured by hexagonal cells mapped to one another by
the translations $\ev_1$ and $\ev_2$ as shown in~(c). The superlattice-two
pattern is left unchanged by $\ttc$, $\kappa_x$ and $\toc\rosixty^3$. The small
circles, dotted lines and shading in (b) and (c) serve to identify equivalent
hexagons related by translations, rotations and reflections of the pattern.}
 \label{fig:sl2box}
 \end{figure}
 \end{center}

\section{Finding the symmetry group~$\Gamma$ and the irrep for the
superlattice-two bifurcation problem}\label{sec:findgamma}

A closer examination of images obtained from the experiment reveals that it is
possible to impose a hexagonal lattice on the observed patterns, whose
instantaneous spatial symmetries are depicted in
{\mbox{figure~\ref{fig:sl2box}}}. The choice of lattice is not unique as it can
be shown that there are many possible candidates (for example a $\sqrt{3}:1$
rectangular lattice), but a hexagonal lattice is a natural choice due to the
symmetry of the standing hexagons. Let us denote the two generating vectors of
the hexagonal lattice $\mathcal{L}$ by $\ev_1, \ev_2 \in \Rset^2$ such that
 \begin{equation}\label{genvect}
 \nom{\ev_1} = \nom{\ev_2} = c, 
 \end{equation}
where $c$ is a scaling factor (figure~\ref{fig:sl2box}(c)). Functions in the
plane that are doubly-periodic with respect to  $\mathcal{L}$ satisfy
 \begin{equation}\label{doublepd}
 u(\xv,t) = u(\xv + \lv,t), \quad \xv = (x,y) \in \Rset^2, 
 \; \lv \in \mathcal{L},
 \end{equation}
where the lattice is defined as
 \begin{displaymath}
 \mathcal{L} = \left\{n_1\ev_1 + n_2\ev_2: \;
 (n_1, n_2) \in \Zset^2\right\}.
 \end{displaymath}

First let us consider the spatial symmetries of the standing hexagons shown
schematically in figure~\ref{fig:sl2box}(a). They are invariant under the
action of $\deesix$ as well as two translations, which we define as follows:
 \begin{equation}\label{tauonetwo}
 \ton: \xv \rightarrow \xv + \tilde{\ev}_1, \quad 
 \ttw: \xv \rightarrow \xv + \tilde{\ev}_2,
 \end{equation}
and let $\nom{\tilde{\ev}_1} = \nom{\tilde{\ev}_2} = c_0$ be the observed size
of the periodic cell in which the basic standing hexagons fit. Our aim is to
pick a value for the scaling factor $c$ in \eqref{genvect} in terms of~$c_0$.
As indicated at the end of section~\ref{sec:groupideas}, a suitable choice of
the value~$c$ is one that gives a hexagonal cell whose size captures precisely
one spatial period of the bifurcating modes, as shown schematically in
figure~\ref{fig:sl2box}(c). In fact, the observed ratio of the two lengths
$\nom{\ev_i}$ and $\nom{{\tilde{\ev}}_i}$ is $c/c_0 = 2\sqrt{3}$, and for this
value of~$c$, the symmetry group~$\Gamma^s$ of the standing hexagons includes
non-trivial translations generated by $\ton$ and~$\ttw$. The structure of the
group $\langle \ton, \ttw \rangle$ can be determined if we express each of the
translations in \eqref{tauonetwo} in terms of $\ev_1$ and~$\ev_2$. We can then
look for the lowest powers $n_1, n_2, n_3, n_4 \in \Zset^+$ such that
$\ton^{n_1}$, $\ttw^{n_2}$, $\ton^{n_3}\ttw^{n_4}$ map the lattice
$\mathcal{L}$ to itself, and thus determine the order of the group
$\langle\ton,\ttw\rangle$.

Guided by the experimental observations, we choose $\tilde{\ev}_1 =
c_0\left(\frac{\sqrt{3}}{2}, \haf\right)$, $\tilde{\ev}_2 = c_0\left(0, 1
\right)$ such that $\ev_1=4\tilde{\ev}_1-2\tilde{\ev}_2=c_0(2\sqrt{3},0)$,
$\ev_2=2\tilde{\ev}_1+2\tilde{\ev}_2=c_0(\sqrt{3},3)$ (see
figure~\ref{fig:sl2box}). The translations can now be written as
 \begin{equation}\label{newtauonetwo}
 \ton : \xv \rightarrow \xv + \frac{1}{6}\ev_1 + \frac{1}{6}\ev_2, \quad
 \ttw : \xv \rightarrow \xv - \frac{1}{6}\ev_1 + \frac{1}{3}\ev_2,
 \end{equation}
and we can easily show that they satisfy $\ton^6 = \ttw^6 = \ton^2\ttw^2 =
\hbox{identity}$, as vectors of the form $\xv + m_1\ev_1 + m_2\ev_2$ for any
integers $m_1$ and $m_2$ lie in $\mathcal{L}$ and are therefore identified.
Since $\ton$ and $\ttw$ commute we can also see that every element generated by
$\ton$ and $\ttw$ can be written as $\ton^n\ttw$ or $\ton^n$ for $n =
0,\ldots\,,5$. In total there are twelve different translations, forming a
group that is isomorphic to $\mrom{Z}_6 \times \mrom{Z}_2$. The order of this
group being twelve corresponds to the fact that each of the large hexagonal
cells in figure~\ref{fig:sl2box}(c) contains exactly twelve of the smaller
hexagons.

\begin{table}[htb]
\begin{center}
\small
\begin{tabular}{|c|r|r|r|r|r|r|r|r|r|r|r|r|r|r|r|} \hline  
  & \multicolumn{15}{c|}{Conjugacy classes of $\Gamma^{s}$} \\
\hline
\multirow{3}{6mm}{Irrep} & $id$           & $\kappa_x$
                         & $\rho$         & $\rho\kappa_x$ 
                         & $\rho^2$       & $\rho^3$     
                         & $\toc\rho^3\kappa_x$ 
                         & $\ttw\rho^2$   & $\toc\rho^3$ 
                         & $\ttw$ 
                         & $\ton\kappa_x$ & $\ttw\kappa_x$ & $\ttc$        
                         & $\toc\kappa_x$ & $\ton^2$ \\
 & & $\ttc\kappa_x$ & & & & & 
     $\ton\ttw\rho^3\kappa_x$ & & 
     $\ton\ttw\rho^3$ & & & & & & \\
 & $(1)\ast$ & $(6)\ast$ & $(24)$ & $(18)$ & $(8)$ 
 & $(3)$ & $(18)\ast$ & $(16)$ & $(9)\ast$ 
 & $(6)$ & $(12)$ & $(12)$ & $(3)\ast$ & $(6)$ & $(2)$ \\ \hline
$\mrom{M}_{\Gamma^s}^{1}$ & $1$ & $1$ & $1$ & $1$ & $1$ 
               & $1$ & $1$ & $1$ & $1$ & $1$ 
               & $1$ & $1$ & $1$ & $1$ & $1$ \\
$\mrom{M}_{\Gamma^s}^{2}$ & $1$ & $-1$ & $1$ & $-1$ & $1$ 
               & $1$ & $-1$ & $1$ & $1$ & $1$ 
               & $-1$ & $-1$ & $1$ & $-1$ & $1$ \\ 
$\mrom{M}_{\Gamma^s}^{3}$ & $1$ & $1$ & $-1$ & $-1$ & $1$ 
               & $-1$ & $-1$ & $1$ & $-1$ & $-1$ 
               & $-1$ & $-1$ & $1$ & $-1$ & $1$ \\
$\mrom{M}_{\Gamma^s}^{4}$ & $1$ & $-1$ & $-1$ & $1$ & $1$ 
               & $-1$ & $1$ & $1$ & $-1$ & $1$ 
               & $-1$ & $-1$ & $1$ & $-1$ & $1$ \\
$\mrom{M}_{\Gamma^s}^{5}$ & $2$ & $0$ & $1$ & $0$ & $-1$ 
               & $-2$ & $0$ & $-1$ & $-2$ & $2$ 
               & $0$ & $0$ & $2$ & $0$ & $2$ \\
$\mrom{M}_{\Gamma^s}^{6}$ & $2$ & $0$ & $-1$ & $0$ & $-1$ 
               & $2$ & $0$ & $-1$ & $2$ & $2$ 
               & $0$ & $0$ & $2$ & $0$ & $2$ \\
$\mrom{M}_{\Gamma^s}^{7}$ & $2$ & $2$ & $0$ & $0$ & $2$ 
               & $0$ & $0$ & $-1$ & $0$ & $-1$ 
               & $-1$ & $-1$ & $2$ & $2$ & $-1$ \\ 
$\mrom{M}_{\Gamma^s}^{8}$ & $2$ & $-2$ & $0$ & $0$ & $2$ 
               & $0$ & $0$ & $-1$ & $0$ & $-1$ 
               & $1$ & $1$ & $2$ & $-2$ & $-1$ \\ 
$\mrom{M}_{\Gamma^s}^{9}$    & $3$ & $1$ & $0$ & $1$ & $0$ 
                  & $3$ & $-1$ & $0$ & $-1$ & $-1$ 
                  & $-1$ & $1$ & $-1$ & $-1$ & $3$ \\ 
$\mrom{M}_{\Gamma^s}^{10}$ & $3$ & $-1$ & $0$ & $1$ & $0$ 
                  & $-3$ & $-1$ & $0$ & $1$ & $-1$ 
                  & $1$ & $-1$ & $-1$ & $1$ & $3$ \\ 
$\mrom{M}_{\Gamma^s}^{11}$ & $3$ & $-1$ & $0$ & $-1$ & $0$ 
                  & $3$ & $1$ & $0$ & $-1$ & $-1$ 
                  & $1$ & $-1$ & $-1$ & $1$ & $3$ \\ 
$\mrom{M}_{\Gamma^s}^{12}$ & $3$ & $1$ & $0$ & $-1$ & $0$ 
                  & $-3$ & $1$ & $0$ & $1$ & $-1$ 
                  & $-1$ & $1$ & $-1$ & $-1$ & $3$ \\
$\mrom{M}_{\Gamma^s}^{13}$ & $4$ & $0$ & $0$ & $0$ & $-2$ 
                  & $0$ & $0$ & $1$ & $0$ & $-2$ 
                  & $0$ & $0$ & $4$ & $0$ & $-2$ \\  
$\mrom{M}_{\Gamma^s}^{14}$ & $6$ & $-2$ & $0$ & $0$ & $0$ 
                  & $0$ & $0$ & $0$ & $0$ & $1$ 
                  & $-1$ & $1$ & $-2$ & $2$ & $-3$ \\  
$\mrom{M}_{\Gamma^s}^{15}$ & $6$ & $2$ & $0$ & $0$ & $0$ 
                  & $0$ & $0$ & $0$ & $0$ & $1$ 
                  & $1$ & $-1$ & $-2$ & $-2$ & $-3$ \\
\hline 
\end{tabular}
\vspace{4mm}
\caption{\sl \small Character table of the group of spatial symmetries
$\Gamma^s$ constructed via the algorithm taken from~\cite{Cornwell}. A
representative element is shown for each conjugacy class, and the number of
elements in the class is given in brackets. Classes marked by $\ast$ contain
elements (specified at the top of the table) of the eight-element group
$\Sigma^{s}_{\zv^{\ast}_q}=\langle\;\ttc,\;\kappa_x,\;\toc\rosixty^3\,\rangle$.
\vspace{8mm}}
 \label{tab:chartable}
 \end{center}
 \end{table}

So in terms of the lattice~$\mathcal{L}$, the full spatial symmetry of the
standing hexagons is given by the group
$\Gamma^s=\left(\mrom{Z}_6\times\mrom{Z}_2\right)\dotplus\deesix$, where
$\deesix$ is generated by a reflection $\kappa_x$ and a $60^{\circ}$~rotation
$\rho$ and its standard action on $\Rset^2$ is given by
 \begin{equation}
 \kappa_x : (x,y) \rightarrow (-x,y), \quad 
 \rho: (x,y)\rightarrow \haf\left(x - \sqrt{3}y, \sqrt{3}x+y\right),
 \end{equation}
and $\mrom{Z}_6 \times \mrom{Z}_2$, an invariant subgroup of~$\Gamma^s$, is
generated by the two translations
 \begin{equation}
 \ton : (x,y) \rightarrow 
              \left(x+ \frac{\sqrt{3}}{2}c_0, y + \frac{1}{2}c_0\right), 
 \quad
 \ttw : (x,y) \rightarrow 
              \left(x, y + c_0\right).
 \end{equation}
The group~$\Gamma^s$ has the semi-direct product structure mentioned in
{\mbox{section~\ref{sec:groupideas}}}. As a result we can apply the algorithm
taken from~\cite{Cornwell} to calculate all its characters and irreps, and we
present the characters of its irreps in {\mbox{table~\ref{tab:chartable}}}.

Any elements that have the same character as the identity in a unitary irrep
of~$\Gamma^s$ must also act like the identity~\cite{Collins90}. Using this
simple idea and the information taken from experimental observations about the
spatial symmetries of the unstable mode, we can single out the irrep of
$\Gamma^s$ that describes the instantaneous symmetry-breaking behaviour.
Careful study of snapshots of the superlattice-two pattern shows that it is
invariant under the action of $\ttc$, $\kappa_x$ and $\toc\rosixty^3$ (see
figures~\ref{fig:original}(b) and~\ref{fig:original}(c), where the pattern is
shown at a slightly tilted angle, and figure~\ref{fig:sl2box}(b)). The group
generated by these elements is by definition the isotropy subgroup of the
bifurcating mode under the action of $\Gamma^s$ (cf~\eqref{isotropy}),
therefore
 \begin{equation}\label{isosub}
 \Sigma^{s}_{\zv^{\ast}_q} \equiv  
 \langle \; \ttc, \;\kappa_x ,\;\toc \rosixty^3\, \rangle
 \subset \Gamma^s.
 \end{equation}
We can now go through the list of characters of $\Gamma^s$ given in
{\mbox{table~\ref{tab:chartable}}} and determine which irrep satisfies the
criteria of permitting $\Sigma^{s}_{\zv^{\ast}_q}$ defined in~\eqref{isosub} to
be an isotropy subgroup of the bifurcating solution. First, any irreps that
satisfy
 \begin{equation}\label{elimcond}
 \chi_{M_{\gamma_s}} = \chi_{M_{id}} \;
 \mbox{for some}\; \gamma_s \in \Gamma^s \;\mbox{and} \;
 \gamma_s \not\in \Sigma^{s}_{\zv^{\ast}_q}
 \end{equation}
must be rejected, because in these cases the isotropy subgroup of
$\zv^{\ast}_q$ must contain spatial symmetry elements apart from those that are
observed. This eliminates representations 1--12. In representation~13, the
class containing~$\ttc$ is represented by the identity, but this class also
contains $\toc$ and $\ton\ttw$, which are not in $\Sigma^{s}_{\zv^{\ast}_q}$,
so eliminating this irrep and leaving only 14 and~15. Then we can use the trace
formula~\cite{GolubitskySS88} to calculate the dimension of the fixed-point
subspace of $\Sigma^{s}_{\zv^{\ast}_q}$:
 \begin{equation*}
 \mbox{dim}\,\mbox{Fix}\left(\Sigma\right) 
 = \frac{1}{\nom{\Sigma}}\sum_{\sigma \in \Sigma} 
 \chi_{M_{\sigma}},
 \end{equation*}
which gives 0 for representation~14 and 1 for representation~15. Clearly we
require $\mbox{dim}\,\mbox{Fix}\left(\Sigma^{s}_{\zv^{\ast}_q}\right)\neq0$,
since $\mbox{Fix}\left(\Sigma^{s}_{\zv^{\ast}_q}\right)$ is non-trivial. Thus
the six-dimensional irrep $\mrom{M}_{\Gamma^s}^{15}$ is the only one in which
$\Sigma^{s}_{\zv^{\ast}_q}$ satisfies the conditions of being an isotropy
subgroup of the observed mode.

In addition to being equivariant under the action of spatial symmetries as
specified by this irrep, the normal form of the period-doubling bifurcation
problem has an extra symmetry corresponding to a translation in time by one
period of the external forcing:
 \begin{equation}\label{timeshift}
 \tau_t: t \rightarrow t + T.
 \end{equation}
This element can be viewed as a spatio-temporal symmetry with a trivial spatial
action, and it acts independently from elements in $\Gamma^s$ with respect to
the standing hexagons. So the full symmetry group~$\Gamma$ of the normal form
for the superlattice bifurcation problem is a direct product between $\Gamma^s$
and the group $\langle \tau_t \rangle$, which, as can be seen
from~\eqref{pdouble} and~\eqref{timeshift}, is isomorphic to~$\mrom{Z}_2$,
hence $\Gamma = \Gamma^s \times \mrom{Z}_2$ as we pointed out in
section~\ref{sec:groupideas}. We can write each element $\gamma \in \Gamma$ as
 \begin{equation}\label{directprod}
 \gamma = \left(\gamma_s, {\sigma}_t\right), \;\gamma_s \in
 \Gamma^s,\;{\sigma}_t \in \langle\tau_t \rangle\, ,
 \end{equation}
such that for $\gamma_1 = \left(\gamma_{s_1}, {\sigma}_{t_1}\right)$, $\gamma_2
= \left(\gamma_{s_2}, {\sigma}_{t_2}\right)$, $\gamma_1\gamma_2 =
\left(\gamma_{s_1}\gamma_{s_2}, {\sigma}_{t_1}{\sigma}_{t_2}\right)$. Because
of the direct product structure of $\Gamma$ and the period-doubling nature of
the bifurcating solution, $\tau_t$ must act like $-1$ on the amplitudes of the
marginal modes. Therefore the irrep of $\Gamma$ that specifies the action of
spatial and spatio-temporal symmetry elements on the marginal modes and the
normal form of $\gv$ can be constructed from the set of matrices
$M_{\gamma_s}\! \in \mrom{M}^{15}_{\Gamma^s}$ as follows:
 \begin{displaymath}
 M_{\gamma} = \left\{ \begin{array}{ll}
                M_{\gamma_s} & \mbox{if} \quad {\sigma}_t = \hbox{identity} \\
               -M_{\gamma_s} & \mbox{if} \quad {\sigma}_t = \tau_t
                \end{array}
              \right.
 \end{displaymath}
for all $\gamma = \left(\gamma_s, {\sigma}_t\right) \in \Gamma$. This irrep,
which we denote by~$\mrom{M}_{\Gamma}$, is of the same dimension as
$\mrom{M}^{15}_{\Gamma^s}$, which implies that we have a six-dimensional centre
manifold at the bifurcation point. So all bifurcating solutions can be written
as $u(\xv, t) = u_0(\xv, t) + \zeta(\xv,t)$ such that
 \begin{equation}\label{planform1}
 \zeta(\xv, qT) = A_qf_1(\xv) + B_qf_2(\xv) + C_qf_3(\xv) + 
 \mbox{c.c.} + \mbox{h.o.t.}, \quad q \in \Zset
 \end{equation}
where $u_0(\xv,t)$ represents standing hexagons, \hbox{c.c.} denotes complex
conjugate, \hbox{h.o.t.} denotes the higher-order terms, and $A_q$, $B_q$,
$C_q\in\Cset$ are the small amplitudes of $f_1$, $f_2$ and $f_3$, the three
complex marginal eigenfunctions that form a basis for the neutral eigenspace
(excluding the two zero eigenvalues corresponding to translating the underlying
pattern). Note that by including the higher-order terms, $\zeta$~represents the
nonlinear perturbation from the standing hexagons.

Applying the method described in~\cite{Cornwell}, we can construct all the
$6\times6$ matrices $M_{\gamma}$ that specify the action of $\Gamma$ on
$\Rset^6$ (or $\Cset^3$) for the irrep~$\mrom{M}_{\Gamma}$. Rather than
describe this procedure, we find it convenient to specify the group action by
choosing a small number of Fourier modes to represent the marginal
eigenfunctions, and working out how the amplitudes of these modes $A_q$, $B_q$
and $C_q$ transform under the generating elements of~$\Gamma$. Since
representations are defined only up to a similarity transformation, the choice
of Fourier modes we make will not matter, as long as we are careful not to
introduce any accidental symmetries (which would become apparent on checking
the characters).

Any function $u(\xv,t)$ defined on the lattice~$\mathcal{L}$ can be written as
a double Fourier series of the form
 \begin{equation}
 u(\xv,t) = \sum_{j_1\in \Zset} \sum_{j_2\in \Zset}
                  u_{j_1,j_2}(t)\, e^{2\pi i \left(j_1\kv_1 + j_2\kv_2\right)
                 \cdot \xv} ,
 \end{equation}
where $\kv_1$ and $\kv_2$ are the generating wavevectors of the dual lattice
$\mathcal{L}^{\ast}$ related to $\ev_1$ and $\ev_2$ by $\kv_i \cdot \ev_j =
\delta_{ij}$ such that \eqref{doublepd} holds. Our choice of the vectors
$\ev_1$ and $\ev_2$ requires $\kv_1 = k\left(\frac{\sqrt{3}}{2},-\haf\right)$,
and $\kv_2 = k(0,1)$, where $k = \frac{1}{3c_0}$.

We use the observed instantaneous symmetry of the pattern (see
figure~\ref{fig:sl2box}(c)) to select a representative function from the full
set of Fourier modes, starting with a single Fourier mode $e^{2\pi
i(j_1\kv_1+j_2\kv_2)\cdot\xv}$ (and its complex conjugate) for some choice of
integers $j_1$ and~$j_2$. If the pattern is to be invariant under $\ttc$,
$j_1$~must be even, so set $j_1=2m$, and, for later convenience, set $j_2=m+n$,
where $m$ and $n$ are integers. With this choice, the Fourier mode is $e^{2\pi
ik(\sqrt{3}mx+ny)}$. The pattern is also invariant under $\toc\rosixty^3$. Now
$\rosixty^3$ replaces the chosen mode by its complex conjugate, and $\toc$
multiplies the mode by a complex number with unit modulus. Since $\toc$ is of
order two but not equal to the identity, it must act by multiplying the mode
by~$-1$. This forces $m+n$ to be odd (and so for the observed pattern, the
amplitude of the Fourier mode must be pure imaginary). The translation $\tau_1$
must act with order~6 (otherwise the pattern would be invariant under a lesser
translation in that direction), so $3m+n\equiv1\mod 6$ or $3m+n\equiv5\mod 6$.
The second of these is essentially the complex conjugate of the first, so we
choose $3m+n\equiv1\mod 6$; $(m,n)$ could be $(0,1)$, $(2,1)$ or $(1,4)$, for
example. Finally, the reflection $\kappa_x$ generates a new function $e^{2\pi
ik(-\sqrt{3}mx+ny)}$, so the superlattice-two pattern can be exemplified by a
mode of the form $f_1=e^{2\pi ik(\sqrt{3}mx+ny)}+e^{2\pi ik(-\sqrt{3}mx+ny)}$.
Sixty degree rotations of this function generate $f_2$ and $f_3$, so we have:
 \begin{equation}\label{marginalmodes} 
 f_1 = e^{2\pi i\Kv_1\cdot \xv} + e^{2\pi i\Kv_2\cdot \xv}, \;
 f_2 = e^{2\pi i\Kv_3\cdot \xv} + e^{2\pi i\Kv_4\cdot \xv}, \;
 f_3 = e^{2\pi i\Kv_5\cdot \xv} + e^{2\pi i\Kv_6\cdot \xv} 
 \end{equation}
(the true eigenfunctions will be made up of linear combinations of such
functions), where
 \begin{center}
 \begin{tabular}{ll}
 $\Kv_1 = k(\sqrt{3}m, n)$,   &
 $\Kv_2 = k(-\sqrt{3}m, n)$,  \\ 
 $\Kv_3 = \frac{k}{2}\left(\sqrt{3}(m+n), (-3m+n)\right)$, &
 $\Kv_4 = \frac{k}{2}\left(\sqrt{3}(-m+n), (3m+n)\right)$, \\
 $\Kv_5 = \frac{k}{2}\left(\sqrt{3}(m+n), (3m-n)\right)$, &
 $\Kv_6 = \frac{k}{2}\left(\sqrt{3}(-m+n), -(3m+n)\right)$.  \\
 \end{tabular}
 \end{center}
These wavevectors have the same wavenumber $K(m,n) = k\sqrt{3m^2+n^2}$, with
$m$ and $n$ satisfying $3m+n \equiv 1\mod 6$. With this choice of basis
functions, the relevant irrep of $\Gamma$ can be specified by the action of the
generating elements of $\Gamma$ on the amplitudes $(A_q, B_q, C_q)$:
 \begin{eqnarray}\label{amplitudeirrep}
 \kappa_x : (A_q, B_q, C_q) &\rightarrow&  (A_q, \bar{C_q}, \bar{B_q}), 
 \label{amplitudeirrepfirst} \\
 \rosixty: (A_q, B_q, C_q) &\rightarrow&  (B_q, C_q, \bar{A_q}), \\
 \ton: (A_q, B_q, C_q) &\rightarrow&  
 (e^{\frac{i\pi}{3}}A_q, e^{\frac{i2\pi}{3}}B_q, e^{\frac{i\pi}{3}}C_q), \\
 \ttw: (A_q, B_q, C_q) &\rightarrow&  
 (e^{\frac{i2\pi}{3}}A_q, e^{\frac{i\pi}{3}}B_q, e^{-\frac{i\pi}{3}}C_q), \\
 \tau_{t}: (A_q, B_q, C_q) &\rightarrow&  (-A_q, -B_q, -C_q). 
 \label{amplitudeirreplast}  
 \end{eqnarray}
We include the subharmonic action of~$\tau_{t}$ here for completeness. The same
representation could be constructed using the method described
in~\cite{Cornwell}.

\section{Normal form of the bifurcation problem}\label{sec:bifprob}

We now have sufficient information to invoke the equivariant branching
lemma~\cite{GolubitskySS88} and describe the different patterns that must be
formed in the instability that created the superlattice-two from standing
hexagons. Before doing this, we will compute the normal form for the
bifurcation since we need it to work out the stability of the various patterns.
The irrep~(\ref{amplitudeirrepfirst}--\ref{amplitudeirreplast}) we identified
in section~\ref{sec:findgamma} implies that the reduced map $\gv$ introduced in
\eqref{normalform} is six-dimensional, and we let ${\mathbf{z}}_q = (A_q, B_q,
C_q)$, $q \in \Zset$, $A_q,B_q,C_q\in\Cset$. As indicated earlier, the action
of $\tau_t$ defined in \eqref{timeshift} is due to the subharmonic nature of
the bifurcating modes with respect to the overall driving period $T$ given in
\eqref{pdouble}. If each iteration in $\zv_q$ corresponds to advancing in time
by $T$, then $\zv_{q+2} = -\zv_{q+1} = \zv_{q}$. Consequently, $\gv(\zv_{q}) =
\zv_{q+1} = -\zv_{q+2} = -\gv(\zv_{q+1})= -\gv(-\zv_{q})$~\cite{ElphickTBCI87}.
So the map $\gv$ will be an odd function of the amplitudes $A_q$, $B_q$ and
$C_q$, as well as being $\Gamma$-equivariant. This information enables us to
write down the form of $\gv$ including up to fifth order terms:
 \begin{eqnarray}
 A_{q+1} &=& -\left(1+\mu\right) A_q + \alpha_1\nomsq{A_q}A_q  + 
           \alpha_2\left(\nomsq{B_q} + \nomsq{C_q}\right)A_q
           + \beta_1\nom{A_q}^4 A_q \nonumber \\
         & & + \mbox{ } 
           \beta_2\left(\nom{B_q}^4 + \nom{C_q}^4\right)A_q 
         + \beta_3\nomsq{A_q}\left(\nomsq{B_q} + \nomsq{C_q}\right)A_q
            + \beta_4 \nomsq{B_q}\nomsq{C_q}A_q \nonumber \\
         & & + \mbox{ }  
                  \beta_5 B_{q}^2 \bar{C}_{q}^{2}\bar{A_q}
                 + \nu \bar{A}_{q}^5 ,
 \label{Aqdot} \\
 B_{q+1} &=& -\left(1+\mu\right) B_q + \alpha_1\nomsq{B_q}B_q  + 
           \alpha_2\left(\nomsq{A_q} + \nomsq{C_q}\right)B_q
           + \beta_1\nom{B_q}^4 B_q \nonumber \\
         & & + \mbox{ }  
            \beta_2\left(\nom{A_q}^4 + \nom{C_q}^4\right)B_q 
         + \beta_3\nomsq{B_q}\left(\nomsq{A_q} + \nomsq{C_q}\right)B_q
            + \beta_4 \nomsq{A_q}\nomsq{C_q}B_q \nonumber \\
         & & + \mbox{ }   
                  \beta_5 A_{q}^2 C_{q}^{2}\bar{B_q}
                 + \nu \bar{B}_{q}^5 ,
 \label{Bqdot}\\
 C_{q+1} &=& -\left(1+\mu\right) C_q + \alpha_1\nomsq{C_q}C_q  + 
           \alpha_2\left(\nomsq{A_q} + \nomsq{B_q}\right)C_q
           + \beta_1\nom{C_q}^4 C_q \nonumber \\
         & & + \mbox{ } 
            \beta_2\left(\nom{A_q}^4 + \nom{B_q}^4\right)C_q 
         + \beta_3\nomsq{C_q}\left(\nomsq{A_q} + \nomsq{B_q}\right)C_q
            + \beta_4 \nomsq{A_q}\nomsq{B_q}C_q  \nonumber \\
         & & + \mbox{ } 
                  \beta_5 \bar{A}_{q}^2 B_{q}^{2}\bar{C_q}
                 + \nu \bar{C}_{q}^5,
 \label{Cqdot}
 \end{eqnarray}
where all coefficients are forced by symmetry to be real.

Apart from the $\nu$~terms, the equations above are equivalent to the
$\mrom{T}^2\dotplus\deesix\times\mrom{Z}_2$-equivariant amplitude equations
(truncated to the same order) that arise in the context of Boussinesq
convection on a hexagonal lattice~\cite{GolubitskySS88,GoluSwif84}, once they
are re-interpreted as amplitude equations rather than a map. The $\nu$~terms
have the effect of breaking the full $\mrom{T}^2$ (two-torus) symmetry group of
translations in a periodic domain to the discrete translations allowed by the
underlying pattern. A natural question to ask is why we needed to work out the
details of the representation before writing down these amplitude equations.
The main reason is that we did not know in advance how many linearly
independent marginal eigenfunctions are involved in the instability. Even if we
had assumed that there were six, it has turned out that there are two
six-dimensional irreps, only one of which is involved in the bifurcation. The
other six-dimensional irrep is generated by taking $f_1=e^{2\pi
ik(\sqrt{3}mx+ny)}-e^{2\pi ik(-\sqrt{3}mx+ny)}$ and (following a similar
analysis) results in the same amplitude equation. Without realising this, one
might conclude incorrectly that patterns that are odd under $\kappa_x$
reflection might also be found in this instability. All the other irreps in
table~\ref{tab:chartable} have dimension less than six (that is, there are
fewer than six independent marginal eigenfunctions), so the order of the
relevant normal forms would be correspondingly less.

We also use~\eqref{hequivariant} to write down the dynamics of the position
$\dv_q$ of the underlying standing hexagons, truncated to quartic order:
 \begin{equation}\label{dqdot}
 \dv_{q+1} = \dv_q + \xi\, \mbox{Im} \left[
                     \begin{array}{c}
                     A_q^2(\bar{C}_q^2 - B_q^2) - 2B_q^2C_q^2 \\
                     -\sqrt{3} A_q^2(B_q^2 + \bar{C_q}^2)
                     \end{array} \right],
 \end{equation}
where $\xi$ is a constant.

\begin{table}[htb]
\begin{center}
\setlength{\extrarowheight}{2pt}
\begin{tabular}{|l|lll|c|}\hline
\multicolumn{3}{|l}{representative solution branch} & 
\multicolumn{2}{l|}{isotropy subgroup \mbox{\hspace{3mm}} 
averaged symmetry}\\\hline
\multirow{2}{6mm}{I} 
 & $1$. 
 & $A\in \Rset$, $B=C=0$ 
 & $\langle\ttc,\;\kappa_x,\;\rosixty^3,\;{\widetilde{\tau}_1}^3\rangle$ 
 & \multirow{2}{28mm}{$\langle\toc,\;\ttc,\;\kappa_x,\;\rosixty^3\rangle$} \\ 
 & $2$.
 & $A\in i\Rset$, $B=C=0$ 
 & $\langle\ttc,\;\kappa_x,\;\toc\rosixty^3,\;{\widetilde{\tau}_1}^3\rangle$ 
 & \\
 \hline
\multirow{2}{6mm}{II}
 & $3$.
 & $A=B=C\in \Rset$
 & $\deesix=\langle\kappa_x,\;\rosixty\rangle$
 & \multirow{2}{28mm}{$\deesix$} \\
 & $4$.
 & $A=-B=C\in i\Rset$
 & $\langle\kappa_x,\;\widetilde{\rosixty}\rangle$
 & \\
 \hline
\multirow{2}{6mm}{III}
 & $5$.
 & $A=0$, $B=C\in \Rset$ 
 & $\langle\kappa_x,\;\rosixty^3,\;{\widetilde{\tau}_2}^3\rangle$
 & \multirow{2}{28mm}{$\langle\ttc,\;\kappa_x,\;\rosixty^3\rangle$}  \\ 
 & $6$.
 & $A=0$, $B=-C\in i\Rset$
 & $\langle\kappa_x,\;\ttc\rosixty^3,\;{\widetilde{\tau}_2}^3\rangle$
 & \\
 \hline
\end{tabular}
\vspace{4mm}
\caption{\sl \small Primary solution branches of the normal form
(\ref{Aqdot}--\ref{Cqdot}) and their isotropy subgroups, grouped into three
types, I--\hbox{III}. Using the analysis presented in
section~\ref{subsec:timeavg} we can show that the two branches within each type
of solution share the same time-averaged spatial symmetries. Branch~2 of type~I
corresponds to the superlattice-two pattern. \vspace{8mm}}
 \label{tab:primarysolns}
 \end{center}
 \end{table}

We can show that there are six isotropy subgroups whose fixed-point subspaces
are one-dimensional, so the equivariant branching lemma tells us that there are
at least six primary bifurcating branches of solutions from standing hexagons,
and we summarise these solutions in table~\ref{tab:primarysolns}. Elements
accented by a tilde represent spatio-temporal symmetries, which, using the
notation introduced in~\eqref{directprod}, can be written as
${\widetilde{\tau}_1}^3 = \left(\toc, \tau_t\right)$, and similarly for
${\widetilde{\tau}_2}^3$ and~$\widetilde{\rosixty}$. The superlattice-two
pattern corresponds to branch~2 of type~\hbox{I}.

\begin{center}
\begin{figure}[htbp]
\begin{tabular}{cccc}
(a) standing hexagons & (b) branch $3$ & 
(c) branch $4$, $t = 0$ & (d)  branch $4$, $t = T$\\
\multicolumn{4}{c}{\resizebox{!}{40mm}{\includegraphics{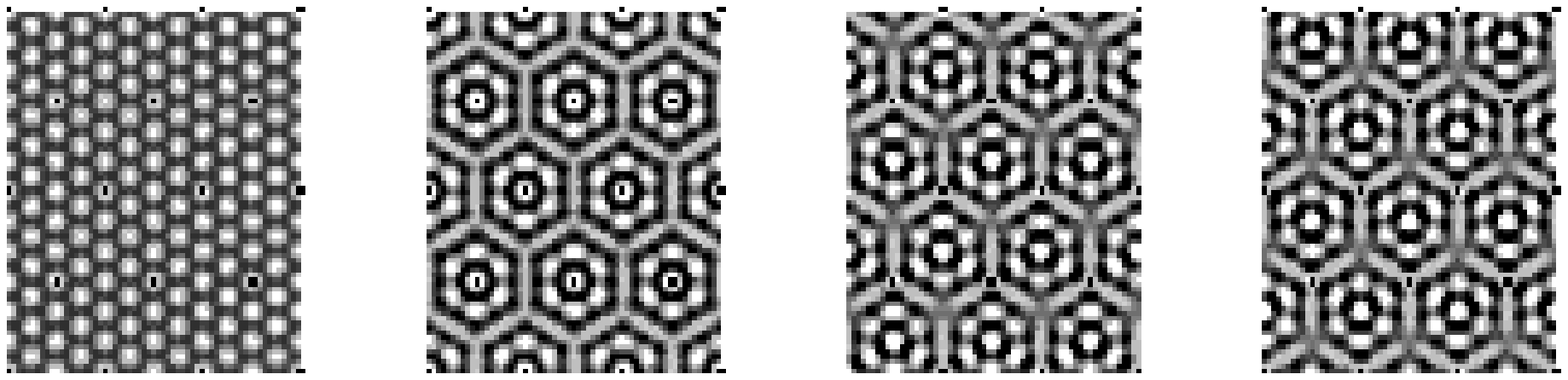}}}\\
(e) branch $2$, $t=0$ & (f) branch $2$, $t=T$& 
(g) branch $1$, $t=0$ & (h) branch $1$, $t=T$ \\
\multicolumn{4}{c}{\resizebox{!}{40mm}{\includegraphics{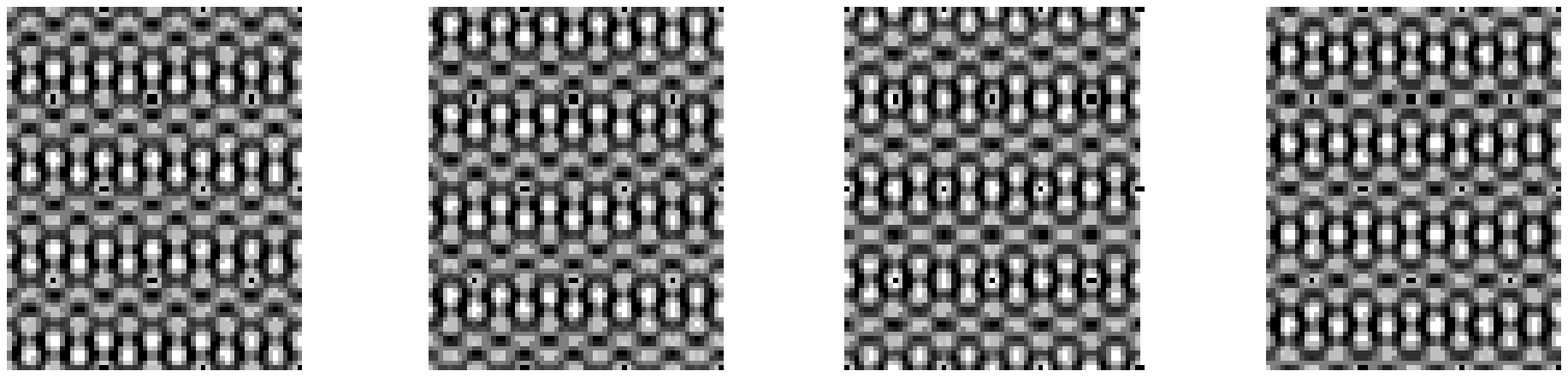}}}\\
(i) branch $5$, $t=0$& (j) branch $5$, $t=T$& 
(k) branch $6$, $t=0$& (l) branch $6$, $t=T$\\
\multicolumn{4}{c}{\resizebox{!}{40mm}{\includegraphics{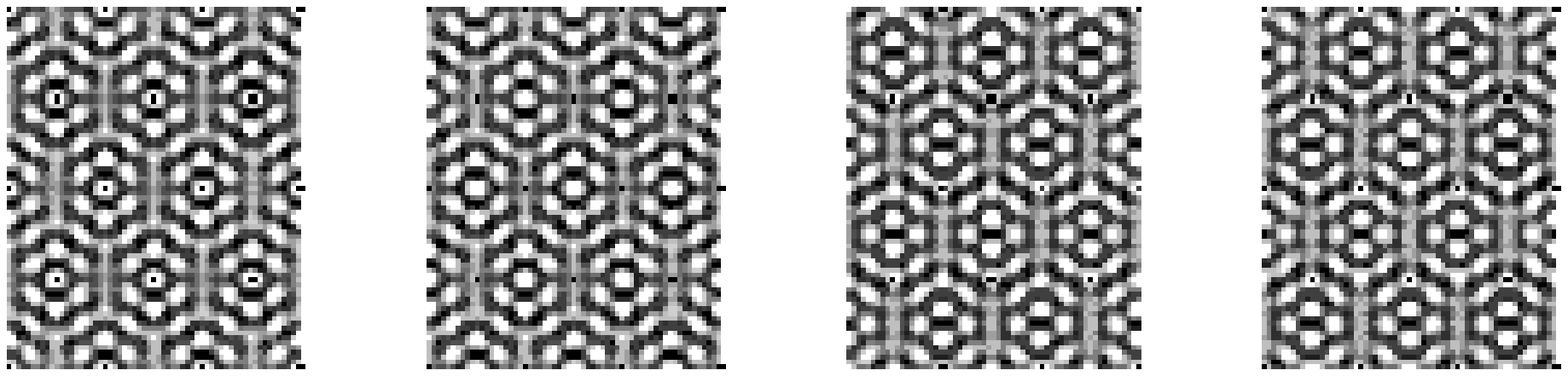}}}
\end{tabular}
\caption{\sl \small Instantaneous planforms of the different solution branches
summarised in table~\ref{tab:primarysolns} and illustrated here in frames
(b)--(l) as small-amplitude perturbations to standing hexagons. Solid squares
represent lattice points of~$\mathcal{L}$.
 (a) Standing hexagons, which have the full $\Gamma$ symmetry.
 (b) Solution branch 3 with $\deesix$ symmetry, referred to
in~\cite{DionSkel97} as `superhexagons'. The periodic hexagonal boxes are
delineated by light borders surrounding each cell.
 (c) \& (d) Solution branch 4 at $t = 0$ and $T$ showing $\mrom{D}_3$ symmetry
as well as the spatio-temporal symmetry~$\widetilde{\rho}$.
 (e) \& (f) The superlattice-two pattern corresponds to solution branch 2 as
they share the same instantaneous spatial symmetries. This pattern is shown
here at $t = 0$ and $T$ with spatio-temporal symmetry
$\widetilde{\tau}_{1}^{3}$ evident.
 (g) \& (h) Similar spatio-temporal symmetry is displayed by solution branch~1.
 (i)--(l) Branches $5$ \& $6$ have very similar symmetry properties: both are
invariant under the action of $\widetilde{\tau}_{2}^{3}$ and instantaneously
they differ only by a shift of the reflection symmetry~$\kappa_y$.}
 \label{fig:solutions}
 \end{figure}
 \end{center}

For the choice of wave integer pair $(m,n)=(2,1)$, the instantaneous planforms
of these six solution branches are illustrated schematically in
{\mbox{figure~\ref{fig:solutions}}. We can compare
figures~\ref{fig:solutions}(e) and~\ref{fig:solutions}(f)
with~\ref{fig:original}(b) and~\ref{fig:original}(c) and notice that the
appearance of stripes at regular intervals in the grey-scale plots of solution
branch 2 closely resembles the essential features of the experimentally
observed superlattice-two pattern.

None of these primary branches leads to a net drift of the underlying hexagonal
pattern; this can be seen in two ways: first, because the rate of drift (from
\eqref{dqdot}, truncated to quartic order) is zero on all six primary branches;
second (and more convincing) since $\rho^3$ is in the symmetry group of all the
time averaged patterns (see below). In other words, the patterns are all pinned
by the $180^{\circ}$ rotation symmetry on average.

\subsection{Stability results}

We summarise in {\mbox{table~\ref{tab:branchevals}}} the branching equations
and the Floquet multipliers of the period~$2T$ patterns, for each of the six
primary solutions guaranteed to exist by the equivariant branching lemma.
Floquet multipliers greater than one in magnitude indicate instability. We
group the six branches into three types and denote them by I, II and III as
shown in tables~\ref{tab:primarysolns} and~\ref{tab:branchevals}. It is evident
that branches within each type are degenerate up to third-order terms, thus
necessitating the inclusion of quintic terms. In particular, only one solution
branch within each of types I and II can be stable depending on the signs
of~$\nu$ and $\beta_5+\nu$, and both branches in type III are always unstable.
Only one branch can bifurcate stably, and all branches must be supercritical
for one of them to be stable. One of the requirements for the observed
superlattice-two pattern (\ie branch 2 of type~I) to be stable is that the
quintic coefficient $\nu>0$. If we also assume the non-degeneracy conditions
$\alpha_1\neq0$, $\alpha_1+2\alpha_2\neq0$, $\alpha_1\pm\alpha_2\neq0$,
$\nu\neq0$ and $\beta_5+\nu\neq0$, close to the bifurcation point the relative
stability of branches of the three types is illustrated by the bifurcation
diagrams shown in figure~\ref{fig:bifndgm}.

\begin{table}[htb]
\begin{center}
\setlength{\extrarowheight}{3pt}
\begin{tabular}{|c|l|l|}\hline
   \multicolumn{2}{|c|}
{Primary solutions and branching equations} & 
Floquet multipliers (multiplicity) \\ \hline \hline
\multirow{4}{4mm}{I} 
&1. $A_q=R_1$, $B_q=C_q=0$, 
 & $1-4\alpha_1 R_1^2$,
   $1-2\left(\alpha_2-\alpha_1\right)R_1^2$ (4 times), \\
& $0 = -\mu + \alpha_1 R_1^2 + \left(\beta_1 + \nu\right)R_1^4$
 & $1+12\nu R_1^4$ \\ 
 \cline{2-3}
&2. $A_q=iR_2$, $B_q=C_q=0$,
 & $1-4\alpha_1 R_2^2$,
   $1-2\left(\alpha_2-\alpha_1\right)R_2^2$ (4 times), \\
& $0 = -\mu + \alpha_1 R_2^2 + \left(\beta_1 - \nu\right)R_2^4$
 &$1-12\nu R_2^4$ \\
 \hline
\multirow{6}{4mm}{II} 
&3. $A_q=B_q=C_q=R_3$, 
 & $1-4\left(\alpha_1+2\alpha_2\right)R_3^2$, \\
& $0 = -\mu + \left(\alpha_1 + 2\alpha_2\right)R_3^2$
 & $1+4\left(\alpha_2-\alpha_1\right)R_3^2$ (2 times), \\
& $\quad{}+\left(\beta_1+2\beta_2+2\beta_3+\beta_4+\beta_5+\nu\right)R_3^4$
 & $1+12\nu R_3^4$ (2 times), 
        $1+12\left(\beta_5+\nu\right)R_3^4$ \\
 \cline{2-3}
&4. $A_q=-B_q=C_q=iR_4$,
 & $1-4\left(\alpha_1+2\alpha_2\right)R_4^2$, \\
& $0 = -\mu + \left(\alpha_1 + 2\alpha_2\right)R_4^2$
 & $1+4\left(\alpha_2-\alpha_1\right)R_4^2$ (2 times), \\
& $\quad{}+\left(\beta_1+2\beta_2+2\beta_3+\beta_4-\beta_5-\nu\right)R_4^4$
 & $1-12\nu R_4^4$ (2 times),
        $1-12\left(\beta_5+\nu\right)R_4^4$ \\
 \hline
\multirow{6}{4mm}{III} &
5. $A_q=0$, $B_q=C_q=R_5$,
 & $1-4\left(\alpha_1+\alpha_2\right)R_5^2$, 
   $1+4\left(\alpha_2-\alpha_1\right)R_5^2$, \\
& $0 = -\mu + \left(\alpha_1 + \alpha_2\right)R_5^2$
 & $1-2\left(\alpha_2-\alpha_1\right)R_5^2$ (2 times), \\
& $\quad{}+\left(\beta_1 + \beta_2 + \beta_3 + \nu\right)R_5^4$
 & $1+12\nu R_5^4$ (2 times) \\
\cline{2-3}
&6. $A_q= 0$, $B_q=-C_q=iR_6$,
 & $1-4\left(\alpha_1+\alpha_2\right)R_6^2$,
   $1+4\left(\alpha_2-\alpha_1\right)R_6^2$, \\
& $0 = -\mu + \left(\alpha_1 + \alpha_2\right)R_6^2$
 & $1-2\left(\alpha_2-\alpha_1\right)R_6^2$ (2 times), \\
& $\quad{}+ 
        \left(\beta_1 + \beta_2 + \beta_3 - \nu\right)R_6^4$
 & $1-12\nu R_6^4$ (2 times) \\ 
 \hline
\end{tabular}
\vspace{4mm}
\caption{\sl \small Branching equations and Floquet multipliers for the six
primary period-two solutions of the normal form~(\ref{Aqdot}--\ref{Cqdot})
listed in table~\ref{tab:primarysolns}. $A_q$, $B_q$ and $C_q$ are complex
amplitudes of the marginal modes defined in~(\ref{planform1}). Only leading
order terms in $R_i$ are shown, and the multiplicities of the Floquet
multipliers (computed from the second iterate of the map) are indicated.
 \vspace{8mm}}
 \label{tab:branchevals}
 \end{center}
 \end{table}
 
\begin{figure}[htb]
\begin{center}
\resizebox{!}{4.0truein}{\includegraphics{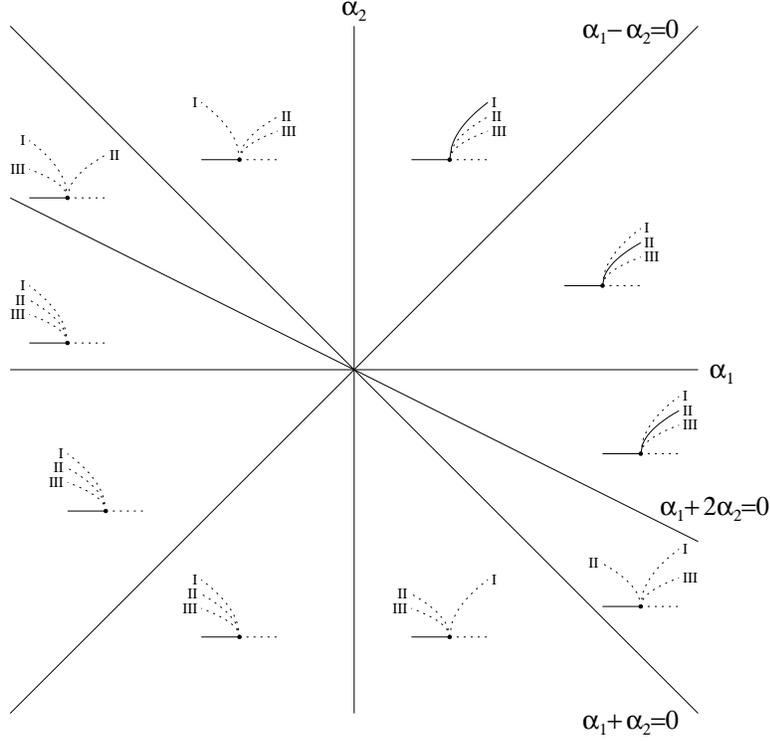}} 
 \caption{\sl \small Bifurcation diagrams for the $\Gamma$-equivariant normal
form (\ref{Aqdot}--\ref{Cqdot}). The sign of the cubic coefficient $\alpha_1$
determines whether solution type~I bifurcates sub- or supercritically. If we
assume that the quintic coefficient $\nu>0$, then the superlattice-two pattern
(branch~2 of type~I) can occur as a stable branch in the region of the
$(\alpha_1,\alpha_2)$ space given by $\left\{(\alpha_1,\alpha_2): \alpha_1>0,\,
\alpha_2>\alpha_1\right\}$.}
 \label{fig:bifndgm}
 \end{center}
 \end{figure}

The experimental results~\cite{KudrPier98} suggest that the bifurcation may
have subcritical branches as there is a parameter regime in which standing
hexagons and the superlattice-two pattern may coexist. On the other hand, the
experimentalists report no hysteresis between standing hexagons and
superlattice-two, while they do report hysteresis between hexagons and other
patterns at other parameter values, so it is not clear whether or not there is
a direct bifurcation from standing hexagons to the superlattice-two pattern in
the experiments. With our parameters, we require $\alpha_1>0$,
$\alpha_2>\alpha_1$ and $\nu>0$ for the superlattice-two pattern (branch~2 of
type~I) to bifurcate stably, but the branch could also be stable in the region
$\alpha_1<0$, $\alpha_2>\alpha_1$ and $\nu>0$ if there were a saddle-node
bifurcation on branch~\hbox{I}.

\subsection{Time-averaged behaviour}\label{subsec:timeavg}

We can study the symmetry properties of the time-averaged image of the observed
solution by integrating over a full period of the newly created periodic orbit
(cf~\cite{BaranyDG93,RuckSilb98,DellnitzGN94}). Specifically, we let
$u_0(\xv,t)$ be the standing hexagons solution and $\zeta(\xv,t)$ the nonlinear
perturbation to this solution such that $u(\xv,t)=u_0(\xv,t)+\zeta(\xv,t)$
represents the observed pattern. We know that $u_0(\xv,t)$ is
$\Gamma$-invariant, \ie $\gamma u_0(\xv,t)=u_0(\xv,t)$ for all
$\gamma\in\Gamma$, and have also found that the spatial and spatio-temporal
symmetry of $\zeta(\xv,t)$ is given by its isotropy subgroup
$\langle\ttc,\;\kappa_x,\;\toc\rosixty^3,\;{\widetilde{\tau}_1}^3\rangle \equiv
\Sigma_{\zeta}\subset\Gamma$. Let $\gamma_s$ and $\gamma_t$ denote respectively
the purely spatial symmetry elements and the spatial part of spatio-temporal
symmetry elements in $\Sigma_{\zeta}$ that act on $\zeta(\xv,t)$ as follows:
 \begin{equation}\nonumber
 \gamma_s \zeta(\xv,t) = \zeta(\xv,t), \quad
 \gamma_t \zeta(\xv,t) = \zeta(\xv,t+T).
 \end{equation}
The time-averaged value of $u(\xv,t)$ can be obtained by integrating over the
full period of the bifurcating solution:
 \begin{eqnarray}
 \bar{u}(\xv) &=& \frac{1}{2T}\int_{0}^{2T} u_0(\xv,t) + 
                 \zeta(\xv,t) \,\ud t 
                  \nonumber \\
              &=& \bar{u}_0(\xv)
                  + \frac{1}{2T}\int_{0}^{T} \zeta(\xv,t) + 
                         \zeta(\xv,t+T) \,\ud t \label{baru}, 
 \end{eqnarray}
where we have used the fact that $u_0(\xv,t) = u_0(\xv,t+T)$. Clearly,
$\bar{u}$ shares the same spatial symmetry with $\zeta$ because both
$\bar{u}_0$ and the individual entries of the integrand in \eqref{baru} are
invariant under the action of $\gamma_s \in \Sigma_{\zeta}$. It is also
invariant under~$\gamma_t$ because the integrand in \eqref{baru} as a whole is
invariant under~$\gamma_t$:
 \begin{eqnarray}
 \gamma_t \bar{u}(\xv) &=& \gamma_t\bar{u}_0(\xv) 
                 + \frac{1}{2T}\int_0^{T} \gamma_t \zeta(\xv,t)
                         + \gamma_t \zeta(\xv,t+T) \, \ud t \nonumber \\
                 &=&     \bar{u}_0(\xv) 
                 + \frac{1}{2T}\int_0^{T} \zeta(\xv,t+T)
                        + \zeta(\xv,t) \, \ud t \nonumber \\
                 &=& \bar{u}(\xv). \nonumber 
 \end{eqnarray}
This result in fact follows readily from more general results on the symmetries
of chaotic attractors~\cite{BaranyDG93,DellnitzGN94}.

In the case of the observed superlattice-two pattern with isotropy
subgroup~$\Sigma_{\zeta}$, the spatial component of the spatio-temporal
symmetry element, namely~$\toc$, will show up alongside $\ttc$, $\kappa_x $ and
$\toc\rho^3$ in the time-averaged image to generate an augmented spatial
symmetry group $\Sigma_{\bar{u}} = \langle \toc, \ttc, \kappa_x, \rho^3
\rangle$. This prediction is in agreement with experimental results
(figure~\ref{fig:original}(a)) and can be understood in the following way. The
action of the translations $\ton$ and $\ttw$ on $\bar{u}(\xv)$ is of order
three since $\toc \bar{u}(\xv) = \ttc \bar{u}(\xv) = \bar{u}(\xv)$, whereas the
order of the same action on $u(\xv)$ is six. So the averaged pattern will
appear to be periodic on a lattice $\mathcal{L}_{\mrom{av}}$ spanned by basis
vectors $\ev_{\mathrm{av}}$ such that $\nom{\ev_{\mathrm{av}}} =
\haf\nom{\ev_i} = \haf c$. We have shown in section~\ref{sec:findgamma} that
$c_0 = \frac{c}{2\sqrt{3}}$, it follows that $\nom{\ev_{\mathrm{av}}} =
\sqrt{3}\nom{\tilde{\ev}_i}$. Therefore the ratio of spatial period of the
averaged pattern to that of the basic standing hexagons is $1:\sqrt{3}$, which
is consistent with the observation reported by~\cite{KudrPier98} as shown in
{\mbox{figure \ref{fig:original}}}(a). Using the same reasoning and information
from the isotropy subgroups of the primary solutions given in
table~\ref{tab:primarysolns}, we therefore predict both branches in each type
of solutions to have the same time-averaged symmetries.

\section{Discussion}\label{sec:discuss}

Starting from the observed instantaneous symmetry of the superlattice-two
pattern reported in~\cite{KudrPier98}, we have been able to show (a)~that a
pattern with the same instantaneous spatial symmetry as the superlattice-two
pattern can bifurcate stably from standing hexagons in a spatial
period-multiplying instability; (b)~that the pattern has the spatio-temporal
symmetry (not reported in~\cite{KudrPier98}) of advancing one driving period in
time combined with a translation by three units in space
(figure~\ref{fig:solutions}(e) and~\ref{fig:solutions}(f)); and (c)~that this
spatio-temporal symmetry accounts for the intermediate spatial scale and
periodicity on a hexagonal lattice of the time-averaged pattern
(figure~\ref{fig:original}(a)). We should emphasise that the intermediate
spatial periodicity of the time-averaged pattern is not the spatial periodicity
of the larger hexagonal lattice that we have assumed.

Arbell \& Fineberg (unpublished) have found the superlattice-two state in their
experiments and have confirmed that it does have the spatio-temporal symmetry
that we predict. Our results also suggest that $60^{\circ}$~rotations are not
in fact symmetries of the time-averaged pattern, but should be weakly broken.
The breaking of $60^{\circ}$~rotational symmetry, if it is present, is
evidently a small effect since the hexagons in figure~\ref{fig:original}(a) do
appear to be invariant under $60^{\circ}$~rotations~\cite{KudrPier98} (this has
been confirmed by Gollub, private communication). For other parameter values,
the symmetry breaking effect may be more pronounced: Fineberg (private
communication) reports that his experimental time-averaged pattern is not
invariant under $60^{\circ}$~rotations. Clearly this would be an interesting
issue to investigate in more detail, but the measurements are delicate and are
liable to be prone to systematic errors or imperfections, so confirming our
prediction could be difficult.

The spatio-temporal symmetry of superlattice-two arises because the instability
of standing hexagons is subharmonic. Other patterns, with different
combinations of spatial and spatio-temporal symmetries, are possible stable
branches in the same bifurcation problem. Not all branches of solutions have
spatio-temporal symmetries, and some of the patterns share the same
time-averaged symmetries even though they have different instantaneous
planforms. The method we have presented is based entirely on symmetry arguments
and is able to deal with instabilities of a fully nonlinear time-periodic
solution.

Spatial period-multiplying instabilities have arisen in a variety of contexts,
in both one~\cite{ProctorWeiss93,Iooss86,AstonSW92,AmdjadiAP97} and two lateral
directions
\cite{KudrPier98,ArbellF98,WagnerMK98,White88,McKenzie88,ProcMatt96,Dawes2000,RWBMP2000}. 
Most of these situations involved relatively simple groups; part of the
difficulty and interest here has been the size of the symmetry group, enlarged
because of the number of translations broken by the new pattern. Only one of
the 15 representations is involved in the superlattice-two bifurcation; other
representations may be relevant to other experiments
(particularly~\cite{ArbellF98,WagnerMK98}) in which standing hexagons lose
stability to patterns that fit into the larger hexagonal cells we have used
here.

As can be seen in section~\ref{sec:findgamma}, a heuristic step in our method
involves the choice of a suitable periodic cell that accommodates the observed
patterns and whose size coincides with exactly one spatial period of the
bifurcating modes. The arrangement of the underlying basic state in this cell
then defines a spatial symmetry group~$\Gamma^s$ of the bifurcation problem and
the instantaneous symmetries of the superlattice instability form its isotropy
subgroup~$\Sigma^{s}$. If a larger hexagonal periodic cell that captures more
than one spatial period of the bifurcating modes had been  chosen, the
translations $\ton$ and $\ttw$ given in~\eqref{tauonetwo} that leave standing
hexagons invariant would have had higher order, resulting in a larger spatial
symmetry group. In this case there would have been more than one irrep of
$\Gamma^s$ in which $\Sigma^{s}$ satisfied the conditions of being an isotropy
subgroup. By choosing the smallest possible periodic cell, we have found that
such indeterminacy can be avoided.

The method we have described in this paper for analysing certain types of
symmetry-breaking instabilities bifurcating from a non-trivial basic state is
based entirely on the observed spatial symmetries of these patterns. However,
information on spatial symmetries of the new pattern alone may not be
sufficient for our approach to be applicable in some problems. For example,
consider a bifurcation problem defined on a spherical domain. Suppose a basic
state with $\mrom{O}(3)$ symmetry loses stability and the observed bifurcating
solutions are axisymmetric, then the isotropy subgroup of the bifurcating modes
is given by~$\mrom{O}(2)$. If the eigenfunctions are expanded in spherical
harmonics, it is known that $\mrom{O}(2)$ is a maximal isotropy subgroup of
$\mrom{O}(3)$ for {\em all} even values of the spherical harmonic
index~$l$~\cite{GolubitskySS88} and so an infinite number of irreps is relevant
to the observed bifurcation. This example illustrates the fact that our method
breaks down if the observed symmetries of the bifurcating modes form an
isotropy subgroup for more than one irrep of the symmetry group of the basic
state.

We are currently involved in applying a similar method to the study of the
`superlattice-one' pattern reported in~\cite{KudrPier98} as a bifurcating
instability from standing hexagons. Preliminary analysis of the experimental
data reveals that a suitable periodic box in this case will give rise to an
arrangement of standing hexagons with a `hidden' reflection
symmetry~\cite{DionSkel97}, which leads to extra complications in determining
the spatial symmetry group. It is an interesting problem that deserves further
investigation.

Unlike some time-periodic solutions (for example, standing rolls), which can
also be defined on a hexagonal lattice, standing hexagons possess only trivial
spatio-temporal symmetries~\cite{RobertsSW86}. So our treatment of the
superlattice patterns as symmetry-breaking instabilities from standing hexagons
is relatively simple because only instantaneous spatial symmetries are needed
to define the isotropy subgroup of these solutions. In general, our approach
can be applied to the study of spatial period-multiplying bifurcations from
solutions with spatio-temporal symmetries and used to investigate some of the
possible symmetry-breaking behaviour, if techniques discussed by Rucklidge \&
Silber~\cite{RuckSilb98} and Lamb~\&~Melbourne~\cite{LambM99} are also
included.

\begin{ack}
We would like to thank Jerry Gollub for inspiring this work, Jonathan Dawes,
Marty Golubitsky, Paul Matthews and Michael Proctor for sharing their insights
with us, and Jay Fineberg for showing us his unpublished experimental results.
DPT is grateful to the Croucher Foundation for financial support. The research
of AMR is supported by EPSRC. The research of RBH was supported by King's
College, Cambridge. The research of MS is supported by NSF grant DMS-9972059,
NSF CAREER award DMS-9502266, and by NASA grant NAG3-2364.

\end{ack}

\bibliographystyle{PRB_normal}
\bibliography{trhs}
   
\end{document}